\documentclass{aastex}
\usepackage{emulateapj5}

\submitted{ApJ Accepted September 24 2001}

\shorttitle{NGC\,985: Making or Breaking a ULIRG?}
\shortauthors{Appleton et al.}

\begin{document}


\title{Mid-IR and CO Observations of the
IR$/$X-Ray Luminous Seyfert I Galaxy NGC\,985: The Making or Breaking of a ULIRG?}

\author{P. N. Appleton\altaffilmark{1}}
\affil{Department of Physics and Astronomy, Iowa State University, Ames, IA 50011}
\email{apple@ipac.caltech.edu}

\author{V. Charmandaris}
\affil{Astronomy Department, Cornell University, 106 Space Sciences Bldg, Ithaca NY 14853}
\email{vassilis@astro.cornell.edu}

\author{Yu Gao}
\affil{Infrared Processing and Analysis Center, MS 100-22, 
California Institute of Technology, Pasadena CA 91125}
\email{gao@ipac.caltech.edu}

\author{F. Combes} 
\affil{Observatoire de Paris, DEMIRM, 61 Ave de l'Observatoire, F-75014 Paris, France}
\email{francoise.combes@obspm.fr}

\author{F. Ghigo}
\affil{NRAO, P. O. Box 2, Green Bank, West Virginia WV 24944}
\email{fghigo@gb.nrao.edu}

\author{C. Horellou}
\affil{Onsala Space Observatory, Centre for Astrophysics \& Space
Science, Chalmers University of Technology, S-43992 Onsala, Sweden}
\email{horellou@oso.chalmers.se}

\author{I. F. Mirabel}
\affil{CEA/DSM/DAPNIA Service d'Astrophysique, F-91191 Gif-sur-Yvette,
France \\ \& Instituto de Astronom\'\i a y F\'\i sica del
Espacio, Conicet, Argentina}
\email{mirabel@discovery.saclay.cea.fr}

\altaffiltext{1}{Now at SIRTF Science Center, California Institute of Technology, Mailcode 249-6, 1200 E. California Av., Pasadena, CA 91125}



\begin{abstract}

We describe ISO\footnote{ISO was an ESA project with instruments
funded by ESA Member States (especially the PI countries: France,
Germany, the Netherlands and the United Kingdom) with the
participation of ISAS and NASA.}  and BIMA observations of the
z\,=\,0.04 Seyfert 1 ring galaxy NGC\,985 which suggest close
parallels with some quasar host galaxies. NGC\,985 contains two
closely spaced nuclei embedded in an R$^{1/4}$--law stellar bulge and
an outer ring, evidence of an ongoing merger. The system contains
$\sim$1.8$\times$$10^{10}$\,M$_{\sun}$ of highly disturbed molecular
gas which lies in an asymmetric bar-like structure with the peak in
observed CO column densities significantly offset from the compact
double nucleus.  In contrast to this, the ISO observations show strong
dust emission centered on the AGN, located in one of the two
nuclei. Fainter CO, MIR, and radio continuum emission provide a glimpse
of the complexities of star formation in the outer ring.

An analysis of the kinematics of the main CO emission reveal
evidence for two dynamically distinct molecular components within
NGC\,985.  The first is a set of isolated super-giant molecular clouds
(SGMCs) which are concentrated within 9--10\,kpc of the active
nucleus.  Although randomly distributed about the center, the clouds
may form part of a clumpy highly-disturbed disk which may be either
just forming around double-nucleus (the making of a ULIRG), or
alternatively may be in the process of being disrupted-perhaps as a
result of a powerful nuclear outflow (the breaking of a ULIRG).  A
second major concentration of CO lies offset from the double-nucleus
in a dynamically coherent ridge of emission in which powerful star
formation is occurring. We tentatively associate CO emission with two
out of six UV absorption-lines seen in the blue wing of the very broad
Ly$\alpha$ emission. Such an association would imply a complex
inter-relationship between the nuclear CO cloud population in
colliding systems, and AGN-driven winds.

\end{abstract}



\section{Introduction}

The detection of large quantities of molecular gas in the host
galaxies of quasars \citep[e.g.][]{Sco93,Schin,guill,Evans01,Pap}
lends strong support to the idea that quasars and molecular-rich
Ultraluminous Infrared Galaxies (ULIRGs) may be related in
evolutionary terms -- an idea first suggested on the basis of their
similar far-infrared (FIR) properties and comparable space densities
\citep{Soifer86,sanders,Evans99}. Such ideas find support in the HST
imaging of nearby quasars, which show that some quasar host galaxies
have disturbed morphologies, and are sometimes ``caught in the act''
of merging \citep[e.g.][]{bah95,bahcall}. Given the potential
importance of the molecular gas (either for quasar fueling, or for the
initiation of major star formation activity in the host), we are
investigating some nearby examples of galaxies which share some
molecular and morphological similarities to host quasars, but are
close enough that detailed observations can be made.

NGC\,985 (Mrk\,1048, VV\,285) is a peculiar galaxy which was first
observed spectroscopically by \citet{DV2}, and contains a Seyfert 1
nucleus in the southern part of the ring. Very broad hydrogen and
\ion{He}{1} $\lambda$5876 lines in excess of 13,000\,km\,s$^{-1}$ are
seen in the nucleus, along with other indications of high excitation
suggesting the existence of an optically-thick accretion disk
\citep{SEM}. The nucleus lies at the center of a large IR-bright
bulge.  Observations at optical and near-IR wavelengths have
demonstrated that this nucleus is double \citep[][ -- see also
archival HST WFPC and NICMOS images]{RES,AM,GPRE} with a separation of
only 3 arcsecs ($\sim$2.5 kpc).  This strongly suggests that the
galaxy is undergoing a merger in which the nucleus of the intruder
galaxy responsible for the ring formation is already sinking towards
the core of the primary galaxy \citep[][]{GPRE}. The one-armed
spiral/ring, and the offset nature of the nucleus that dominates the
luminosity from the radio through to the X-rays, suggest some
similarities with some recently imaged quasar hosts galaxies. In
particular, the quasars PKS\,2349-014 and PKS\,0316-346
\citep{bah95,bahcall}, have very similar morphologies to NGC\,985,
containing highly offset nuclei and outer ring loops. Although milder
examples of offset nuclei are common in samples of collisional ring
galaxies, offsets as extreme as the one in NGC\,985 are very rare
\citep[see][~for a review]{asm96}.

NGC\,985 has an IR luminosity of L$_{\rm
IR}\,=\,1.8\times10^{11}$\,L$_{\sun}$, which is the largest in a
sample of ring galaxies observed by IRAS \citep{asm87}, placing it in
the category of LIRGs \citep{sanders96}.  The absolute magnitude of
NGC\,985 \citep[M$_V$~=~-22.58 from][]{opt_rings} places it in a
similar range of luminosities to that of radio-loud quasar hosts
\citep{bahcall}, despite the fact that NGC\,985 contains a radio
source of only a few mJy \citep[this paper -- see also][]{ulv84}.
NGC\,985 is a powerful X-ray source \citep{Brandt} with a spectrum which is
consistent with both a hard power-law component and a substantial
soft component. The X-ray spectrum is best fitted if an additional
warm-absorber is placed along the line of sight to the AGN.

Recent HST STIS observations (S. Penton \& J. Stocke, U. of
Colorado: Private Communication) also indicate the presence of at
least 6 separate narrow UV absorption lines in the blue wing of the
broad Ly$\alpha$ emission line. We will argue in this paper that two
of these lines may have CO emission counterparts, suggesting a
possible interaction between a nuclear AGN-driven outflow and the
molecular gas. 

The primary nucleus \citep[hereafer N1--following the terminology
of][]{GPRE} of NGC\,985 has a systemic velocity of
12,814$\pm$15\,km\,s$^{-1}$ (z~=~0.04) based on absorption-line
studies performed by \citet{arribas}. The secondary nucleus (N2) has a
systemic velocity of 13,096$\pm$15\,km\,s$^{-1}$.  Assuming a Hubble
constant of 75\,km\,s$^{-1}$\,Mpc$^{-1}$, we adopt throughout this
paper a distance to NGC\,985 of 170\,Mpc, so in our figures an angular
distance of 10 arcsec corresponds to a projected distance of
$\sim$8\,kpc.

Details on our observations are presented in Section 2 and the main
imaging results are shown in Section 3. In Sections 4 and 5 we
elaborate on the kinematics and dynamics of the galaxy based on the
CO, while in Sections 6 and 7 we present the MIR evidence for
peculiarities in the star formation properties of the nucleus. In
Section 8 we discuss the physical conditions in the outer ring
\ion{H}{2} regions, and in Section 9 we explore a potential link with
of the observed UV absorption lines. In Section 10 we discuss the
current and future evolution of NGC\,985, and our conclusions are
presented in Section 11.

\section{Observations}\label{sect_obs}
\subsection{Space-based Mid-IR Observations} 
 
NGC\,985 was observed by ISOCAM, a 32$\times$32 pixel array on board
the ISO satellite \citep{Cesarsky,Kessler} on December 31 1997 (ISO
revolution 776) as part of an ISO-GO observation of bright northern
ring galaxies \citep[see
also][]{cart_iso,7zw_iso,iso_rings,vc_rings}. The galaxy was imaged
through 4 filters, LW1 (4.5~ [4.0-5.0]\,$\mu$m), LW7 (9.62
[8.5-10.7]\,$\mu$m), LW8(11.4 [10.7-12.0]\,$\mu$m), and LW3(15.0
[12.0-18.0]\,$\mu$m) with on-source time of $\sim$10 minutes per filter.
 
A lens resulting in a 3 arcsec pixel field of view, was used to create
a 3$\times$3 raster map in a ``microscan'' mode in both directions
(the telescope was moved 4 arcsecs between each element of the raster
in a regular pattern), and the full width at half maximum of the
point-spread-function (PSF) was $\sim$4 arcsecs at 12\,$\mu$m. The
overall field of view was 104$\times$104 arcsecs$^2$, which easily
encompassed the whole galaxy which is over 40 arcsecs across.
 
The data were analyzed with the CAM Interactive Analysis software
(CIA)\footnote{CIA is a joint development by the ESA astrophysics
division and the ISOCAM consortium}. A dark model taking into account
the observing time parameters was subtracted. Cosmic ray hits were
removed by applying a wavelet transform method \citep{Starck}.
Corrections of detector memory effects were done applying the
Fouks-Schubert's method \citep{Coulais}. The sky was subtracted using
the emission from areas of the map well outside the galaxy and its
value was within the one expected from the coarse DIRBE zodiacal
maps. The flat field correction was performed using the library of
calibration data.  Finally, individual exposures were combined using
shift techniques in order to correct the effect of jittering due to
the satellite motions (amplitude $\sim$1 arcsec). Based on our
experience with ISOCAM data reduction and application of the above
methods to a large set of observations, we estimate that the
absolute uncertainty in our photometry measurements is less than 20$\%$.

\subsection{Ground-based Radio and Millimeter Observations}

Observations were made with the D-array of the VLA on August 25 1992
at 4.9 ($\lambda$6) and 8.4 GHz ($\lambda$3.5\,cm). These data were
flux and phase calibrated, bad data were flagged and ignored, and
final maps were made using the AIPS routine IMAGR. The effects of
sidelobes were removed using the CLEAN algorithm, resulting in images
of NGC\,985 with synthesized beam sizes of 19.25$\times$13.23 arcsecs
($\lambda$6\,cm) and 10.67$\times$7.55~arcsecs ($\lambda$3.5\,cm).

The $^{12}$CO(1-0) observations of NGC\,985 were carried out with the
Berkeley-Illinois-Maryland Association (BIMA) 10-element millimeter
array \citep{welch} at Hat Creek, northern California.  The
preliminary BIMA observations were carried with one full track
(7--8\,hrs) of C array and D array observations in April 18 1998 and
June 22 1998, respectively.  A nearby phase calibration source
0238+166 was observed for several minutes before and after every
half-hour of observations of NGC\,985. Further observations with both
the C array in May 21--24 1999 (three 6hr tracks and one short 3\,hr
track), and the D array in July 15 and 17 1999 (two 3--4\,hrs tracks).
The source 0339-017 was observed as a phase reference instead of
0238+166 during this observing because it was closer to NGC\,985, and
was bright enough to be used at this time (the source is variable).
Individual antenna system-temperatures averaged 350\,K, and were
typically within the range 250 to 550\,K.  Mars and/or Uranus was
observed for flux calibration.  These data were flux and phase
calibrated, and bad data flagged and channel maps made using the
MIRIAD software.  The synthesized beam of the final maps was
11.94$\times$6.81 arcsecs with a position angle of -1 degrees. The
receivers were tuned to 110.50 GHz corresponding to a redshift of
z~=~0.043, and the correlator was configured to cover a total velocity
range of just over 1500\,km\,s$^{-1}$ at a spectral resolution of
$\sim8$\,km\,s$^{-1}$(3.1 MHz).

A data-cube containing 40 channel maps was made, each channel covering
a velocity width and separation of 25\,km\,s$^{-1}$, centered on the
heliocentric velocity of 12,934\,km\,s$^{-1}$, and so cover a velocity
range of almost 1000\,km\,s$^{-1}$. Those maps with bright emission
were CLEANed to remove the effects of sidelobes. Subsequent analysis
of these data was performed in AIPS and MIRIAD.

\subsection{Astrometry}

As the next section shows, a large offset is found between the peak in
the map of CO column densities, and the position of the AGN as defined by
observations at radio and IR wavelengths. Because our observations
were made under excellent conditions over seven different tracks with
two arrays, BIMA was expected to return exceptionally high-precision
astrometry. However to confirm this, we constructed maps of the galaxy
by splitting the observations into two separate independent data-sets
which used different combinations of phase calibrators and array
configurations. The results showed identical emission features with
excellent positional and flux agreement. We are therefore confident
that the BIMA astrometry is of very high quality.

Similar arguments can be made for the VLA data which has subarcsec
accuracy, and the AGN core was independently observed by
\citet{ulv84}.  Later in the paper we also compare our observations
with UKIRT K-band IR images published previously by \citet{AM}. These
deep observations were registered to the astrometrically accurate
2MASS K-band image of NGC\,985\footnote{We also independently checked
the astrometry in the 2MASS image by comparing the position of a
foreground star in the northern portion of the image with an
astrometric measurement made in the optical by \citep{jeske}. They
showed perfect agreement.} while the details of the registration of
the MIR images is discussed in the next section.  We are therefore
confident that the registration of all the various datasets used in
the forthcoming discussion are accurate to an arcsecond or better --
an angular scale much smaller than the observed offset in the CO
distribution -- the result that prompted us to perform these checks.

\section{Imaging Results at Different Wavelengths}\label{sect_morphology}

\subsection{Visible, IR and Radio Continuum Comparisons}

\begin{figure} 
\figurenum{1} 
\epsscale{0.8}
\plotone{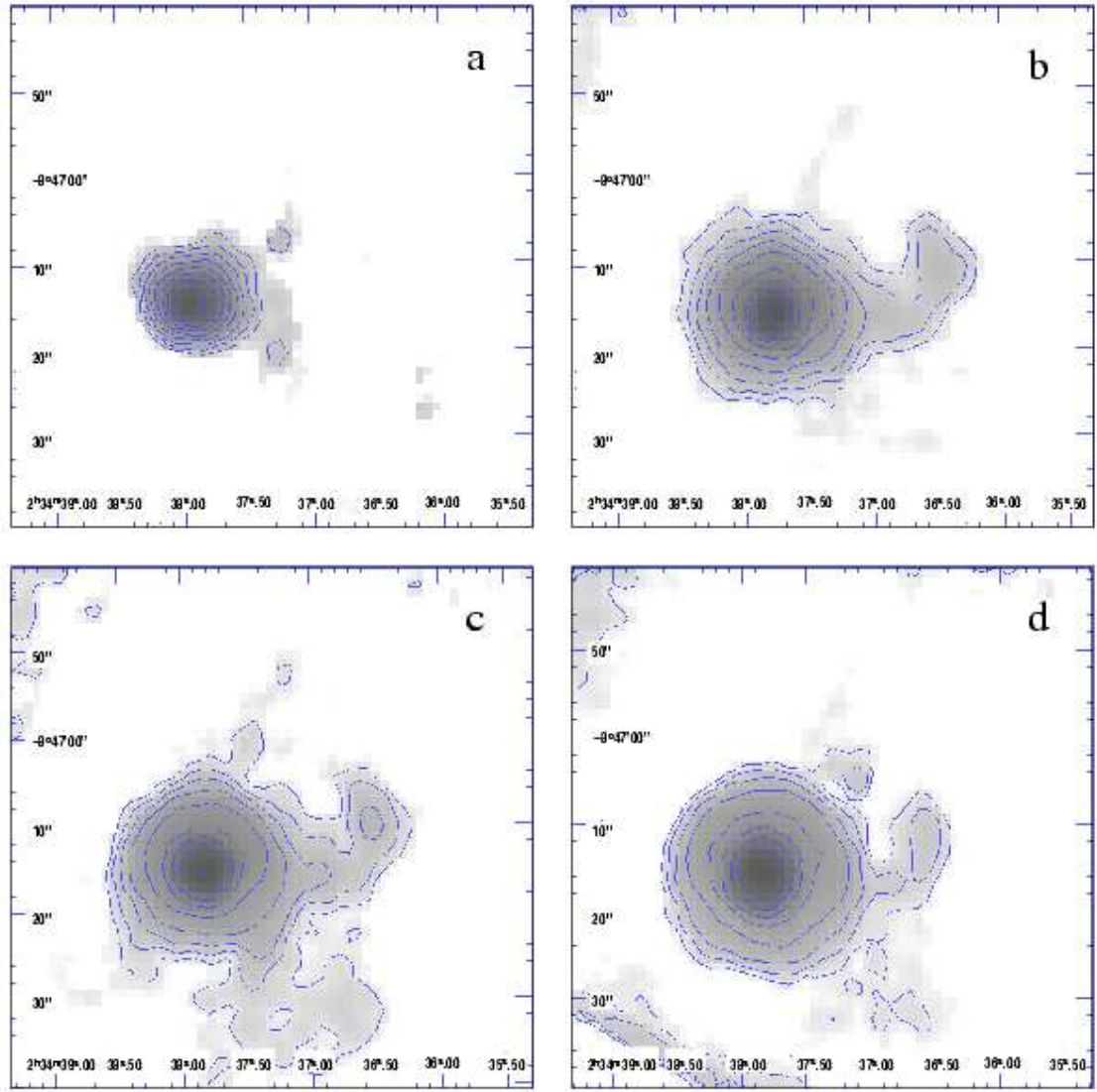}
\caption{ a) The ISOCAM 5\,$\mu$m image of NGC\,985, with an
overlay of nine of its own contours spanned equally logarithmically
between 0.04 and 1.56 mJy\,pixel$^{-1}$. The coordinates are in J2000.
b-c-d) Same as for a) but for the 9.62\,$\mu$m, 11.4\,$\mu$m,
and 15\,$\mu$m image of the galaxy respectively. The spacing of
the nine contours of each image is set as in a) but with limits [0.02
-- 2.47] mJy\,pixel$^{-1}$ for the 9.62$\mu$m, [0.02 -- 3.1]
mJy\,pixel$^{-1}$ for the 11.4$\mu$m and [0.02 -- 3.4] mJy\,pixel$^{-1}$ for
the 15$\mu$m respectively.  \label{isomap}}
\end{figure}

\begin{figure} 
\plotone{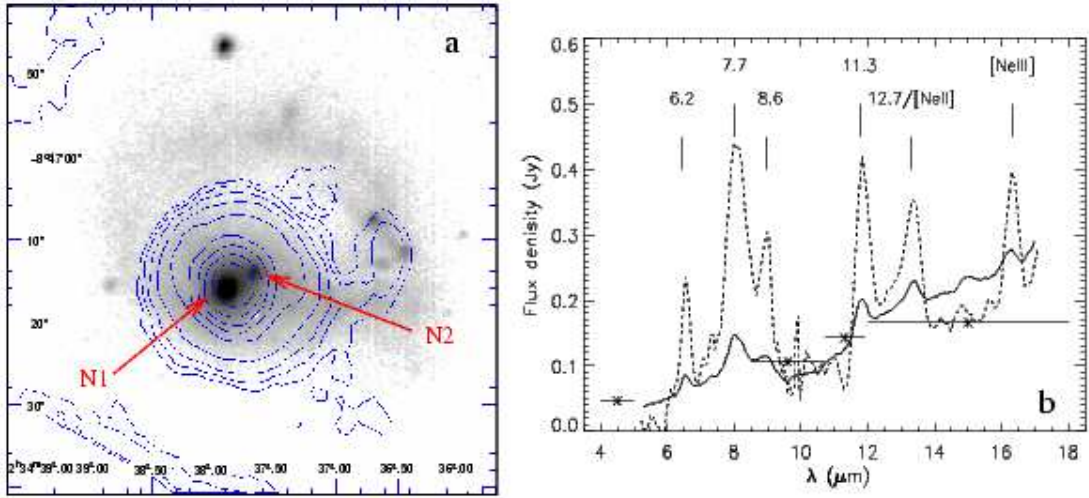}
\figurenum{2} 
\caption{ a) A contour map of the 15\,$\mu$m image shown in Fig.~\ref{isomap}d)
superimposed on an optical R-band image of NGC\,985 kindly provided by
A. M. P\'erez Garc\'{\i}a and J. M. Rodr\'{\i}guez Espinosa (IAC,
Spain). The positions of the two nuclei N1 and N2 are marked. b) The
nuclear spectrum of NGC\,985 as measured with a circular aperture 12
arcsec in radius. The width of each broad-band filter used is marked
with a horizontal line. For comparison we display the spectrum of
NGC\,1068 \citep[from Fig.3 of][]{emeric} in solid line, and that of
the star forming knot B of the Antennae galaxies
\citep[see][]{vigroux} in dashed line. Both spectra have been
red-shifted to the distance of NGC\,985 and normalized to the flux
density of NGC\,985 ISOCAM LW8 (11.4$\mu$m) filter.\label{spectra}}
\end{figure}

The images obtained with ISOCAM at wavelengths of 4.5, 9.62, 11.4 and
15\,$\mu$m respectively are displayed in Figure~\ref{isomap}a-d. For
reference, Figure~\ref{spectra}a shows a contour map of the 15\,$\mu$m
emission superimposed on a grey-scale R-band image of the galaxy. We
also present in Figure~\ref{spectra}b the MIR spectral energy
distribution (SED) of the nucleus, the properties of which are
discussed in detail in Section~\ref{sect_mirsed}. The 15\,$\mu$m
filter is the broadest and has the best signal to noise ratio of the
four filters used. The images obtained in the three of the longer
wavelength filters (Fig.~\ref{isomap}b, c, and d) show very similar
MIR distributions.  The strongest emission comes from the nucleus of
the galaxy and considerable extended emission is present. Emission is
also seen near the two optically prominent
\ion{H}{2} regions in the western part of the outer ring. In addition
fainter emission is detected immediately adjacent to the nucleus to
the west, especially in Figure~\ref{isomap}c and d along the optically
prominent emission ridge.  Only the center of the galaxy is detected
at 4.5\,$\mu$m, but observations made with this filter are less
sensitive than those made at other wavelengths.

Table~\ref{tbl_mir} lists the MIR fluxes of the four ISOCAM filters
for the central bulge/nucleus and for the bright ``knot'' seen near
the western \ion{H}{2} regions in the ring. Additional MIR
observations of NGC\,985 from the ground had also been performed by
\citet{maiolino} who found that the 10.6\,$\mu$m (N-band) flux
density within a 5.3 arcsec aperture centered on the nucleus of
NGC\,985 is 143 mJy.  This agrees to better than 10\% with the average
flux density we obtained at with the 9.7 and 11.4\,$\mu$m ISOCAM
filters, and is conclusive proof that the majority of the 10\,$\mu$m
flux from NGC\,985 is very centrally concentrated, since the ISO flux
was determined over a much larger aperture than the ground-based
observation and yet the same flux was observed.  This is very
different from the situation at 2.2\,$\mu$m, where the AGN's
contribution is much smaller than that of the stellar bulge (see
Section~\ref{sect_kinematics}) in the inner regions.

\begin{figure} 
\plotone{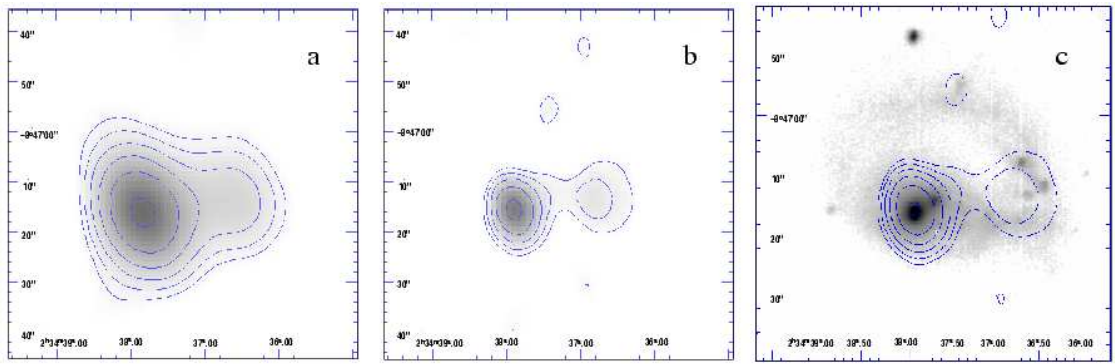}
\figurenum{3} 
\caption{ a) The VLA 6\,cm (C-band) radio continuum map of NGC\,985,
with an overlay of its own contours. The contours are 0.3, 0.5, 0.8,
1.2, 1.9, and 3.0 mJy\,beam$^{-1}$. b) Same in a) but for the 3.5\,cm
(X-band) radio continuum. The contours are 0.2, 0.3, 0.5, 0.8, 1.3,
and 2.0 mJy\,beam$^{-1}$. c) Same contours as shown in b) but
overlayed on the R-band image of NGC\,985 shown in Fig. 2a. \label{radiomap}}
\end{figure}

In order to register the MIR maps with the optical, near-IR, and radio
maps, we could not assume that the absolute astrometry of ISO was
better than a few arcsecs (unlike the relative astrometry which is
much more accurate). We have made the reasonable assumption that the
bright peak in the 15\,$\mu$m map corresponds to the Seyfert nucleus
detected at radio wavelengths.  We registered this peak to the
corresponding peak found from the VLA radio
maps. Figures~\ref{radiomap}a and b show the radio continuum maps
corresponding to a wavelength of 6 and 3.5\,cm respectively.
Figure~\ref{radiomap}c shows the 3.5\,cm map superimposed on the
R-band image of the galaxy. The 3.5\,cm map, which has the better
spatial resolution, shows a bright radio source we associated with the
Seyfert nucleus, and a second component which lies to the west just
inside the outer ring. The radio fluxes are given in
Table~\ref{tbl_mir}.  We fitted a Gaussian to the peaks in both radio
maps (using the AIPS routine IMFIT) to determine an accurate position
for the nucleus. Both positions agreed within a fraction of an
arcsec. The centroid in both maps was found to occur at $\alpha({\rm
J2000}) = 02\fh34\fm37.74\fs~(\pm0.01\fs)$ and $\delta({\rm J2000}) =
-08\fdg47\farcm15.0\farcs~(\pm0.27\farcs)$.

One potential difficulty of using the position of the radio core for
the ISO image registration is the possibility that the nucleus might
be heavily obscured, even at MIR wavelengths,
\citep[e.g.][]{rig,soifer,soifer2}, and the peak may not be spatially
coincident with the radio nucleus. However we can exclude this
possibility because we find excellent positional agreement between the
radio nucleus (above) and a K-band image of the nucleus obtained from
the astrometrically accurate 2MASS survey \citep{2mass}. Since the
position of the near-IR nucleus and the radio nucleus correspond, we
can safely assume that obscuration is not a factor in determining the
position of the AGN core at wavelengths longward of 2\,$\mu$m.

\begin{figure} 
\plotone{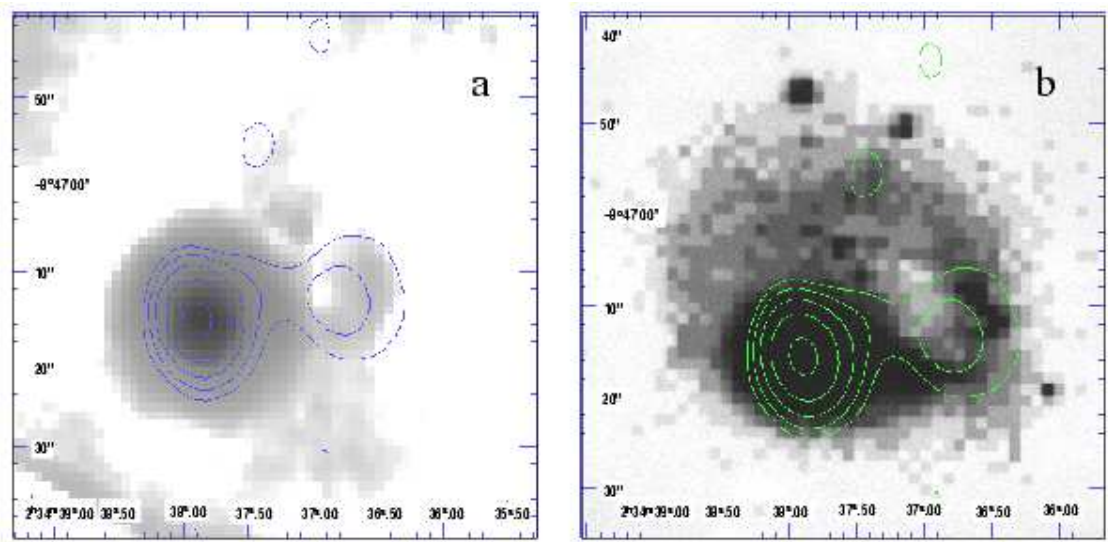}
\figurenum{4} 
\caption{a) The 3.5\,cm (X-band) contours of NGC\,985 overlayed on the
15\,$\mu$m image of the galaxy. The contour spacing is the same as in
Fig~\ref{radiomap}b.  b) Same overlay contours as in a) but this time
on the UKIRT K-band image of NGC\,985 ~\citep[see Figures 1 and 2
in][]{AM}. \label{xband}}
\end{figure}

Figure~\ref{xband} displays the 3.5\,cm radio map superimposed on
images made at 15\,$\mu$m with ISOCAM (Fig.~\ref{xband}a), and at
2.2\,$\mu$m with UKIRT \citep[Fig.~\ref{xband}b, an image based on
near-IR data from ][]{AM}. This shows that the MIR and radio emission
have similar distributions. Figure~\ref{xband}b shows that the radio
source is associated with the eastern (brighter) component of the double
nucleus. No obvious radio emission is found associated with the second
nucleus, consistent with the suggestion by \citet{GPRE} that the
second nucleus is not active\footnote{P. N. Appleton, in a personal
communication, mistakenly reported to J.M. Rodr\'{\i}guez Espinosa
that the second nucleus was also a radio source, a report that was
included in their paper on the double nucleus \citep{GPRE}.  Although
it is true that the radio source appears double, the second component
is seen on a much larger scale and is not associated with the second
nucleus (Figure~\ref{radiomap}c). The mistake is regretable.}.

\subsection{CO Distribution and Integrated Properties of NGC\,985}\label{sect_co}

\begin{figure} 
\plotone{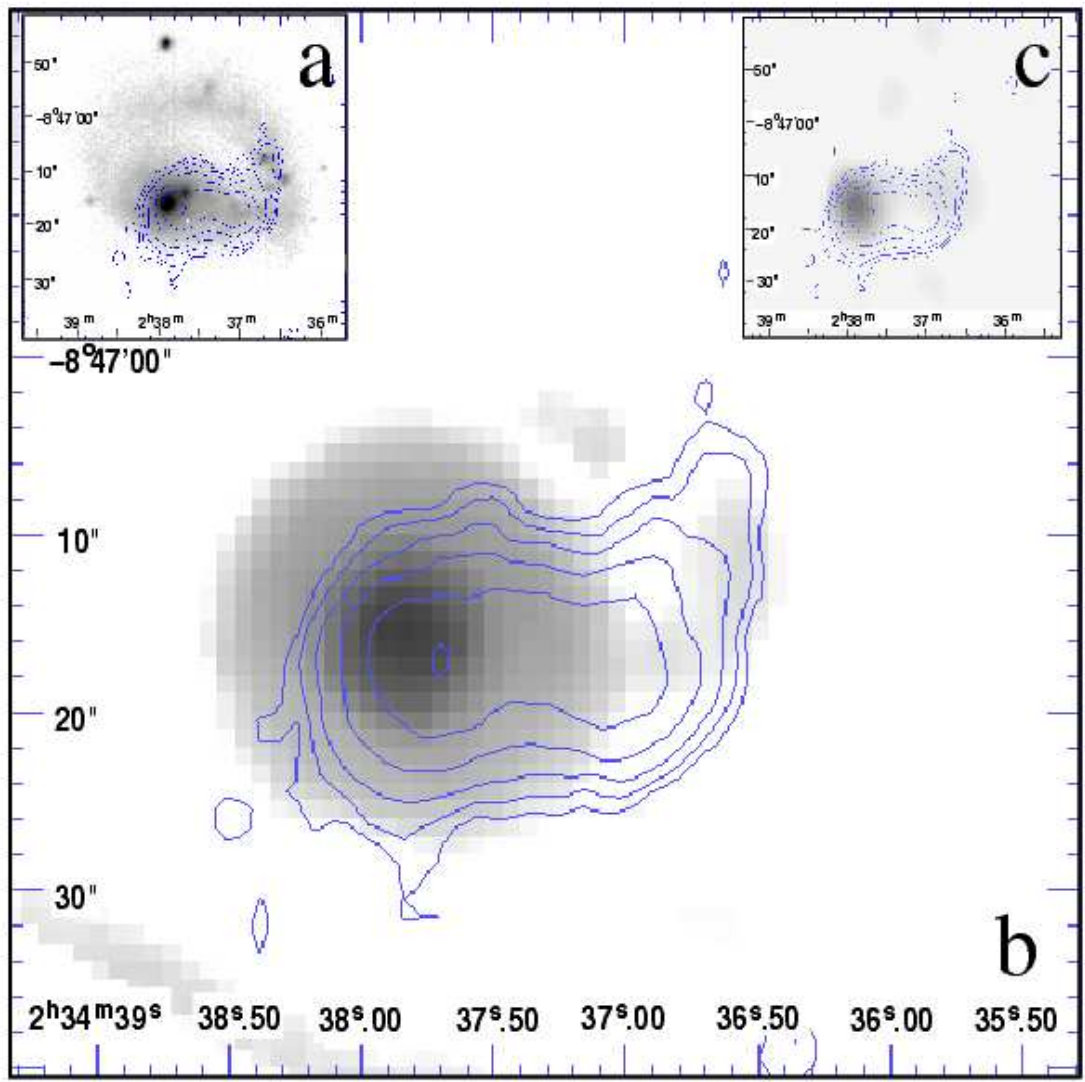}
\figurenum{5} 
\caption{
a) The integrated $^{12}$CO(1-0) emission in contours, over the
optical R-band image of NGC\,985. The contour levels are 4.0, 5.2, 6.8,
8.8, 11.5 and 15.0 Jy\,beam$^{-1}$\,km\,s$^{-1}$.  b) Same as in a)
but the background is the 15\,$\mu$m image of the galaxy.  c) Same as
in a) but the background is the 3.5\,cm (X-band) radio continuum map.
\label{comap}}
\end{figure}

Figure~\ref{comap} shows a map of the integrated $^{12}$CO(1-0)
emission from NGC\,985 superimposed on optical (Fig.~\ref{comap}a),
15\,$\mu$m MIR (Fig.~\ref{comap}b), and 3.5\,cm radio maps
(Fig.~\ref{comap}c) of the galaxy. The emission is distributed in an
elongated structure with two almost equal peaks separated by 13--15
arcsecs.  A weaker third CO concentration is seen to the west, near a
part of the ring which contains two powerful \ion{H}{2} regions. No
emission is seen associated with the large stellar ring which loops to
the north of the nucleus/budge at large radii.

A striking offset of about 5 arcsecs is found between the main body of
CO emission, and the bright radio/IR peak. Indeed the majority of the
CO avoids the nucleus and is mainly concentrated to the west, where
some of it is associated with the peculiar ridge of optical/IR
emission extending away from the center.  Such offsets are unusual in
collisional galaxies, where either the gas is centered on (or
straddles) the nucleus. In the unusual case of Arp\,220, a galaxy with
two nuclei probably undergoing a major merger, the CO is mainly
concentrated between the two nuclei in a massive molecular disk
\citep{Sco97}, but is still centrally concentrated.

Very few ring galaxies have been observed in CO with high
resolution. In a few cases, single dish observations have provided
some indication of molecular distribution on the scale of the large
outer rings. The major single-dish study of ring galaxies by
\citet{cathy} shows that some systems (e.g. II\,Hz\,4) have emission
concentrated in the outer star-forming rings \citep[see also][~for
AM0644-741]{higdon}. However, only two systems, Arp\,118 and Arp\,143,
have been published with high spatial resolution
\citep{gao,higdon97}. In Arp\,118 most of the CO is found in a
powerful star forming ring, with little CO(1-0) emission from the
center of the galaxy. In contrast to this, Arp\,143 \citep{higdon97}
and the Cartwheel \citep{cart-co} have CO concentrated towards the
center.

\begin{figure} 
\plotone{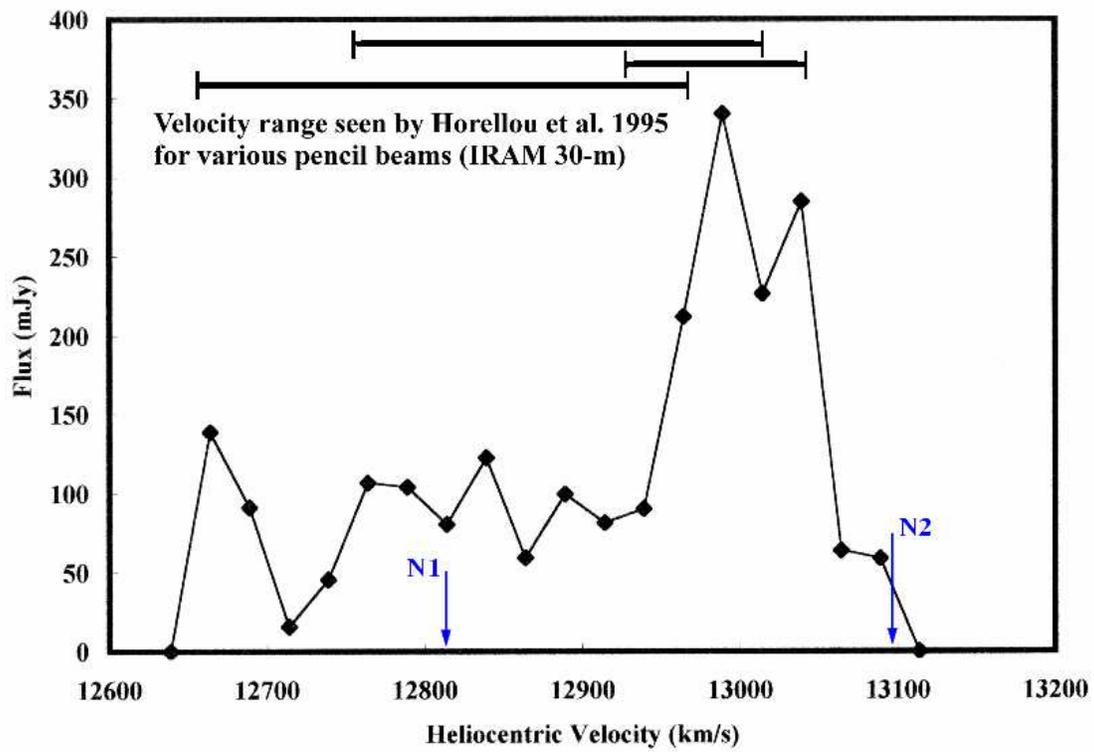}
\figurenum{6} 
\caption{Integrated spectrum of the $^{12}$CO(1-0) emission from our BIMA
data. The optical velocities of the two nuclei, N1 and N2, as well as
the velocity ranges seen by the single dish observations of
\citet{cathy} are marked.\label{cospectrum}}
\end{figure}

We have constructed an integrated spectrum of the CO emission from the
galaxy which is shown in Figure~\ref{cospectrum}, along with arrows
indicating the velocities of the two nuclei. Also shown is the range
of velocities quoted in the single-dish study of NGC\,985 by
\citet{cathy}.  This underestimates the full range observed, since it
shows the range at FWHP for some representative spectra (an integrated
spectrum is not provided in that work).  Their IRAM 30-m $^{12}$CO(1-0)
observations, obtained with a 22\,arcsec beam and performed over
several pointings across the galaxy, agree closely in terms of
overall velocity spread and line-shape with the BIMA spectrum.  There
is also excellant agreement between the BIMA fluxes and the IRAM
observations.  

Table~\ref{tbl_global} presents a compilation of the observed and
derived molecular gas properties based on these observations, along
with published single dish CO and upper limits to the \ion{H}{1} mass
in the system from other work. As the table shows, the BIMA
observations retrieve 100\% of the single dish flux. Adopting the
``standard'' conversion from CO flux to H$_2$ mass (see \citet{young}) 
given by
M(H$_2$)\,=\,1.18$\times$10$^4$\,F$_{\rm CO}$\,D$^2$\,(1+z)$^{-1}$,
with D~=~170\,Mpc, then M(H$_2$) is found to be
1.8$\times$10$^{10}$~M$_{\sun}$. This conversion factor is highly
uncertain, but we adopt a standard conversion for comparison with
other work. The ratio of L$_{\rm IR}$/M(H$_2$) for NGC\,985 is 10 (see
Table~\ref{tbl_global}).  This indicates that the present star
formation activity in NGC\,985 is comparable with a local starburst
galaxy, but not as as high as typical ULIRGs or quasars
\citep{sanders96,Evans01}. From a star formation point of view, NGC\,985 has
the potential to become significantly more luminous and more star
formation efficient in the future.

We note that no neutral hydrogen has yet been detected from NGC\,985.
An upper limit to the \ion{H}{1} mass of $<$
3.6$\times$10$^9$\,M$_{\sun}$ is derived from \citet{hec} after
correcting their upper limit for our adopted value of the Hubble
constant, and assuming that the \ion{H}{1} velocity-width is similar
to that of the CO profile. This leads to a M(H$_{2}$)/M(\ion{H}{1})
ratio of greater than 5.5 implying that {\em the bulk of the cool
interstellar medium (ISM) in NGC\,985 is in molecular
form}\footnote{Contrary to NGC\,985, the Seyfert galaxy
NGC\,1144\,(=Arp\,118) \citet{B99} showed the \ion{H}{1} and CO
distribution and kinematics in that galaxy were highly peculiar as a
result of a violent interaction. Even in that case, the observed
M(H$_{2}$)/M(\ion{H}{1}) ratio was 2.6 -- a factor of at least two
lower than in this case.}.  \citet{felix} have shown that
M(H$_{2}$)/(M\ion{H}{1}) ratios as large as 4-20 are usually found in
advanced merger galaxies that have the highest far-infrared to optical
flux ratios, are ultraluminous in the infrared, and exhibit OH
megamaser emission. In this context, if the upper limit for the HI
mass is confirmed, NGC 985 could be considered as an ultraluminous
infrared galaxy in the making. We will discuss this and another
possibility later in the paper.

\section{CO Kinematics -- Evidence for Two Dynamically Distinct Populations}\label{sect_kinematics}

One reason that the main concentration of CO emission is not centered
on the nucleus of NGC\,985 may have to do with the peculiar motions
present in the CO gas. To demonstrate this, we have constructed a data
cube which represents the observed CO column densities in each of 40
channels covering a velocity spread of almost 1000\,km\,s$^{-1}$
centered on the velocity of the system.

\begin{figure} 
\plotone{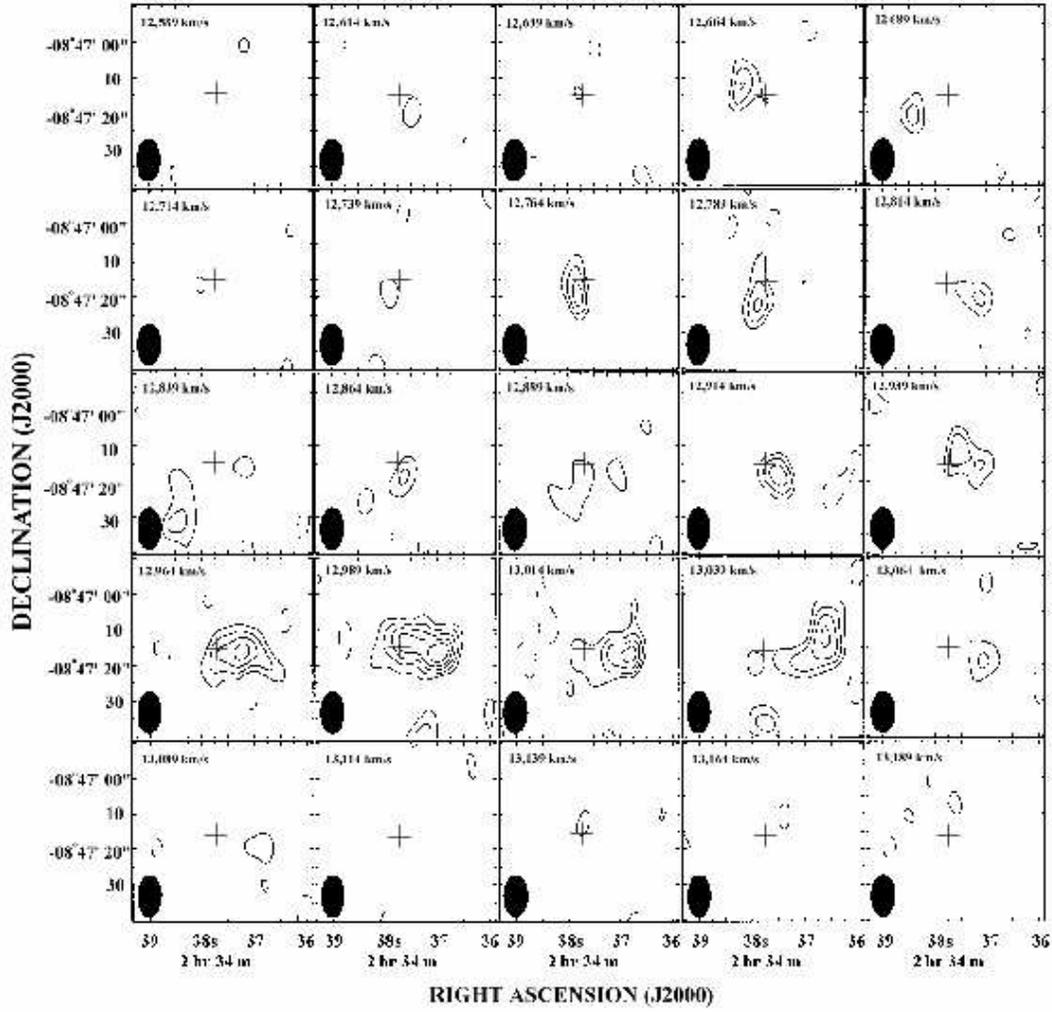}
\figurenum{7}
\caption{
Channel maps of the $^{12}$CO(1-0) emission from NGC\,985. The
velocity of each channel is marked on the upper left corner. The
contour levels are n$\times$0.015\,Jy\,beam$^{-1}$ where
n=-2,2,3,4,5, and 6. Negative contours are marked with dashed
lines. The first positive contour is close to the 3$\sigma$ noise level. 
\label{chanmap}}
\end{figure}

Figure~\ref{chanmap} shows a series of channel maps over the range of
velocities 12,639\,km\,s$^{-1}$ to 13,089\,km\,s$^{-1}$ in steps of
25\,km\,s$^{-1}$ intervals--the range in which significant CO emission
is detected. We note that only a subset of the full range of velocity
coverage is shown here, since the rest of the channels show little
significant emission above an rms noise of 11\,mJy\,beam$^{-1}$. The
first contour is close to the 3$\sigma$ level (30\,mJy\,beam$^{-1}$).

\begin{figure} 
\plotone{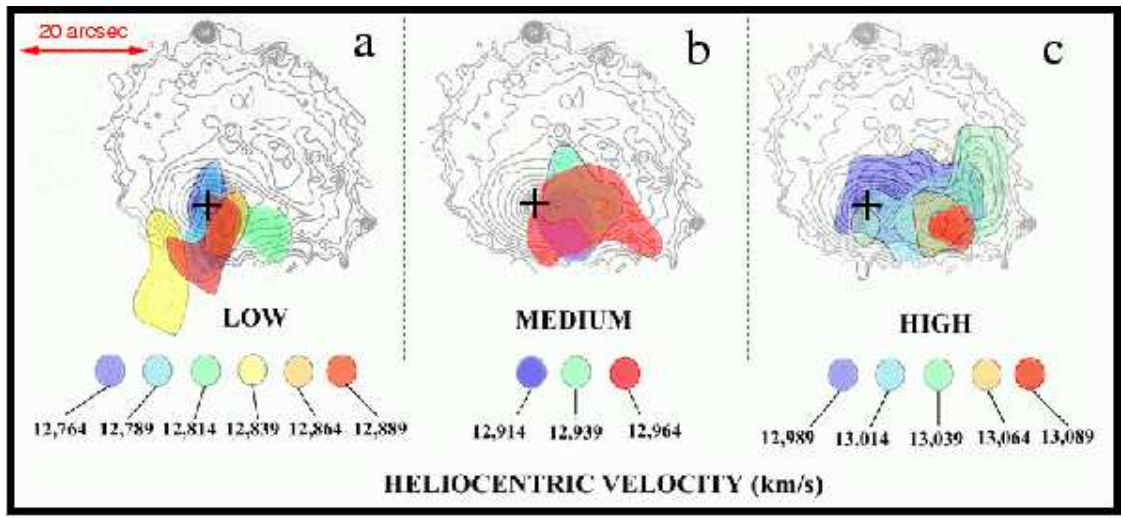}
\figurenum{8} 
\caption{A composite of the various emission features superimposed
on the K-band UKIRT image of the galaxy. Because the kinematics of the
galaxy is unusually complex we have split the figure into three parts,
representing the low velocity, medium velocity and high velocity
regimes in the data cube (a, b \& c respectively). The scale in the
upper left corner indicates the angular extend of all the features
displayed (see Section~\ref{sect_kinematics}).\label{ukirt}}
\end{figure}

To aid the discussion, we also constructed a composite of the various
emission features from Figure~\ref{chanmap} and superimposed onto the
K-band UKIRT contours of the galaxy.  Because the kinematics of the
galaxy is unusually complex, we have split the composites into three
parts, representing the low, medium, and high velocity regimes in the
data cube (Fig.~\ref{ukirt}a, b \& c respectively). In the following
paragraphs we will demonstrate that in Fig.~\ref{ukirt}a, b, the
concentrations of CO are somewhat randomly distributed about the
nucleus showing very little obvious coherence from one channel to the
next, whereas, Fig.~\ref{ukirt}c shows emission that is much more
easily comprehended, as it displays very clear channel-to-channel
coherence. This is not a signal-to-noise effect, since all the
features in the maps are significantly above the noise, and are
identifiable as emission in the single-dish observations. Later, we
will demonstrate that the emission features seen in the first two
panels of Fig.~\ref{ukirt} may have distinctly different dynamical
properties from the more coherent high-velocity emission.

The lowest velocity channel maps in which CO is detected are shown in
Figure~\ref{ukirt}a, covering the range 12,764 to
12,889\,km\,s$^{-1}$. At the these velocities, gas is seen covering
the eastern AGN (N1), but shifts away, first to the south
(V~=12,789\,km\,s$^{-1}$), and then to the west at velocities higher
than 12,814\,km\,s$^{-1}$. Indeed at velocities close to the systemic
velocity of N1 \citep[12,814\,km\,s$^{-1}$ see][]{arribas} the CO
emission is offset from N1 by a few arcsecs and does not cover either
N1 or the western nucleus N2. The tendency of the brighter emission to
be offset to the west of the AGN is the reason why the peak in the
integrated map in Figure~\ref{comap} is offset from N1.

In Figure~\ref{ukirt}b we show the gas at intermediate
velocities. Here again, as in Fig.~\ref{ukirt}a, emission is seen
which has little obvious spatial coherence from one channel to the
next. In some channels. the gas is roughly associated with the ridge
of starlight seen by \citet{RES} extending from the nucleus to the
west. We note that in one channel (V~=~12,939\,km\,s$^{-1}$) there is
a component of the CO seen to the northwest of the AGN. This
corresponds to the position of the possible MIR arc discussed later in
Section~\ref{sect_asymetry}, although this could be coincidental.
However, for the most part, the intermediate velocity gas is only
loosely associated with the linear optical ridge which dominates the
western side of NGC\,985.

The high velocity emission from NGC\,985 begins to show more noticable
channel-to-channel coherence. This is one of the features that
distinguishes it from the lower-velocity emission.  At
V~=~12,989\,km\,s$^{-1}$, gas is seen over a wide range of positions,
extending all the way from the western extreme edge of the ring to the
nucleus N1 (this also shows that the CO emission spectrum in the
direction of the nucleus contains multiple components). Emission over
successively higher velocities is seen to shift progressively towards
the outer ring, bending northwards (at V~=~13,039\,km\,s$^{-1}$) into
the region of two powerful \ion{H}{2} regions. In these channels, a
fainter ``finger'' of emission extends back towards the
double-nucleus.  At extremely high velocities the emission is
concentrated at the end of the stellar ridge. The highest velocity
gas, and that with the highest velocity dispersion (150\,km\,s$^{-1}$)
is seen approximately 14 arcsecs west of the nucleus at the position
$\alpha({\rm J2000})=02\fh34\fm36.9\fs$, $\delta({\rm
J2000})=-08\fdg47\farcm18\farcs$. This position corresponds to a
region in the linear optical emission ridge which exhibits the
strongest H$\alpha$ emission as discussed by \citet{RES} and
\citet{AM}.

The molecular gas distribution and kinematics at first seem quite
complicated and difficult to understand. However, some insight into
the behavior of the clouds comes when we realize that there may be two
different kinds of dynamics being exhibited here: a) cloud-complexes
that are distributed at random positions close to the nucleus and have
no obvious optical/IR counterpart, and b) cloud structures that have
well-defined spatial-velocity coherence. These latter clouds tend to
be associated with the obvious optical/IR features, and look similar
to the structures we are familiar with in studying gas in galaxies
(i.e. spiral arms, rings, tidal loops etc).

\begin{figure}
\plotone{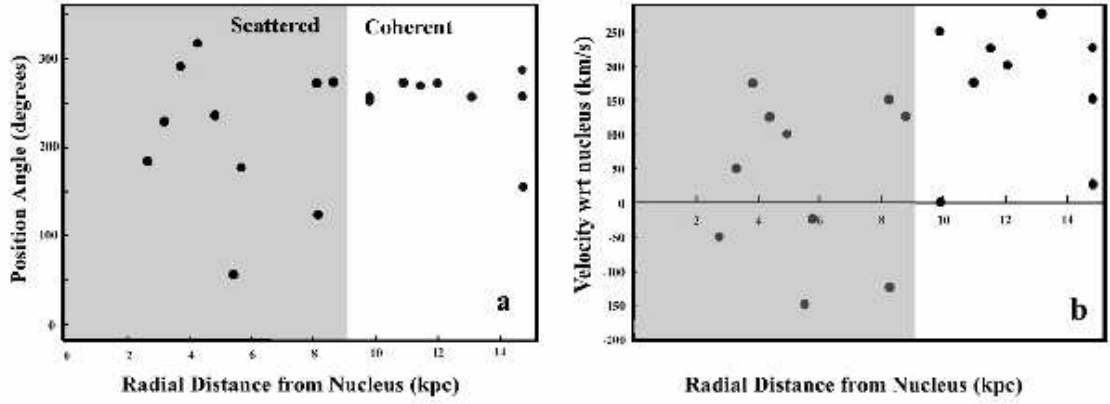}
\figurenum{9} 
\caption{
a) The position angle as a function of distance from the nucleus N1 of
the scattered and coherent SGMCs observed within 15\,kpc. b) Same as
in a) but plotting the velocity of the SGMCs with respect to N1 as a
function of distance. Note how at larger radii the clouds cluster
together in one region of each diagram, a result of spatial and
kinematic coherence (see discussion in
Section~\ref{sect_kinematics}).\label{blobs}}
\end{figure}

Many of the individual ``clouds'' we refer to here are quite massive,
a few $\times$$10^{8}\,M_{\sun}$, and we will refer to them as Super Giant
Molecular Clouds (SGMCs). To show that two dynamically different CO
emission features may be present, we show in Figure~\ref{blobs} two
ways of visualizing them.  Figure~\ref{blobs}a shows the distribution
of position angles of the SGMCs (north through east) relative to the
AGN (nucleus N1) as a function of the radial distance from N1. These
are the coordinates of the centroids of each SGMC measured
channel-by-channel from the channel maps. The centroid of each SGMC
was determined (to within 1--2 arcsecs = 0.8--1.6\,kpc) if it was
represented by two or more contour levels ($>$4.5\,$\sigma$) in the
channel maps of Fig.~\ref{chanmap}. In some cases where extended
emission is observed, only the brighter obvious clumps were extracted.
However, in most cases the identification of a SGMC feature was
straightforward and unambiguous. From Figure~\ref{blobs} it is obvious
that within a radial distance of 9\,kpc (10.9 arcsecs) of the center,
the SGMC complexes are distributed in a random manner with respect to
the nucleus, whereas, the average nature of the emission changes and
becomes more ordered and coherent at larger radii. This transition was
noted in the discussion of the previous figure, but is put on a more
quantifiable footing here.

The separation of the SGMCs into the two components is not restricted
to their different spatial distributions, but also is apparent in
their different kinematic behavior. In Fig.~\ref{blobs}b, we plot the
velocity of the SGMCs (relative to the AGN systemic velocity of
12,814\,km\,s$^{-1}$) as a function of radial distance from the
nucleus. It can be seen that within 9\,kpc (representing the area of
the ``randomly distributed'' SGMCs) the velocities of the clouds are
distributed equally on either side of the systemic velocity of the
active nucleus N1, and the clouds appear to dynamically related to the
nucleus. However, for the emission at larger radii (mainly those
associated with the optical/IR ridge feature) the clouds cluster
together in one region of the diagram, a result of spatial and
kinematic coherence.

The implication of Fig.~\ref{blobs} is that the CO emission in
NGC\,985 exhibits two distinct kinds of dynamics. A randomly distributed
(radius $<$ 9--10\,kpc) SGMC system, centered both spatially and in
velocity-space on the nucleus N1, and a more distant coherent SGMC system
which is highly correlated with obvious optical/IR galaxy components. 

\subsection{Is the ``random component'' rotating?}\label{random}

It is common when analyzing data cubes containing spectral line data
to create ``moment'' maps (isovelocity and velocity dispersion) which
are often useful in helping the astronomer interpret their data.
Although we constructed such maps early in our analysis of these data,
we felt the maps were misleading because, i) the data cube contains
many places where multiple velocity components are seen along the line
of sight and, ii) the cube contains isolated cloud components which
are not obviously related to another. This stochasticity in the
velocity field leads to a very confused pattern in the isovelocity
map, making it difficult to interpret.  Despite these reservations,
{\em some limited} inference can be drawn if we divide it up into two
velocity regimes.
 
\begin{figure}
\plotone{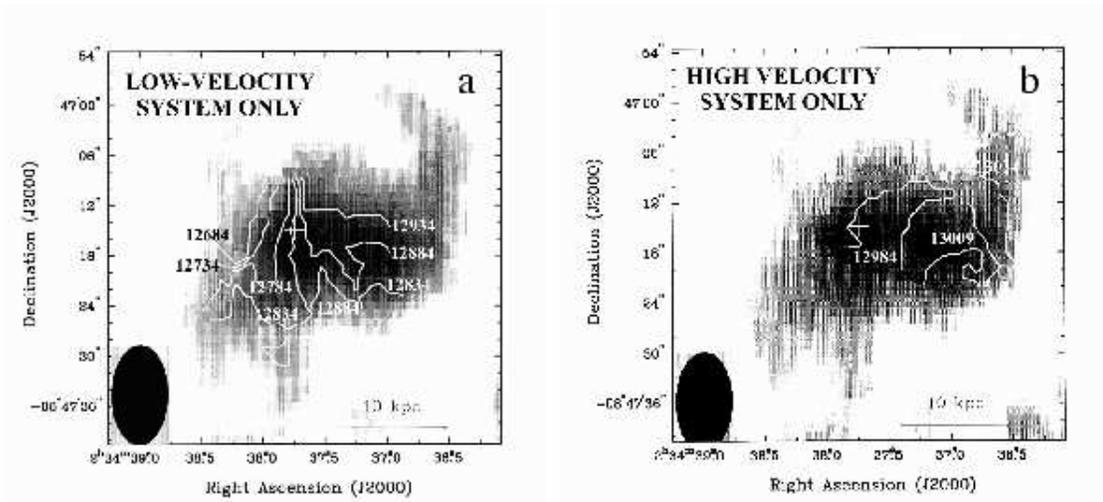}
\figurenum{10} 
\caption{
a) The isovelocity field of the gas for the low-velocity system
superimposed on a grey-scale representation of the CO integrated
column density map--heliocentric velocities are in km\,s$^{-1}$. b)
same as a but for the high velocity system (see text in
Section~\ref{random}).\label{mom1}}
\end{figure}

In Fig.\ref{mom1}a and b we show the isovelocity maps for the cube
covering the velocity range 12,684 to 12,934\,km\,s$^{-1}$, and 12,984
to 13,034\,km\,s$^{-1}$ respectively.  This separation roughly
corresponds to the CO emission associated with the randomly
distributed nuclear-centered clouds (Fig.\ref{mom1}a), and
higher-velocity more coherent clouds (Fig.\ref{mom1}b) discussed
earlier\footnote{We stress that even this simplistic division of the
cube obscures some of the details that are best seen in
Fig~\ref{ukirt}. For example, the isovelocity contour at
12,934\,km\,s$^{-1}$ in Fig.\ref{mom1}a does not smoothly join in a
coherent way with the contours at 12,984\,km\,s$^{-1}$ shown in
Fig.\ref{mom1}b -- a result of multiple components of emission along
several lines of sight.}. Despite some major excursions from
regularity, Fig.\ref{mom1}a shows that the isovelocity contours for
the nuclear-centered low-velocity system may show the signature of
rotation or gas streaming motions, even though they are observed as
discrete clouds with somewhat random {\em positions} relative to the
nucleus.  The higher velocity emission is associated with mainly
identifiable optical/IR features in the outer galaxy and can be seen
in Fig.\ref{mom1}b. Its pattern confirms our earlier observation that
the gas in the higher velocity channels behaves differently from the
low-velocity emission -- the steepest velocity gradient being associated
with the outer eastern ring.

\section{The Dynamics of the Low-velocity CO Clouds and the Ionized
Inner Disk of NGC\,985}\label{dynamics}

In the previous section we presented evidence that the CO emission
seen within 10 kpc of the double nucleus of NGC 985 is composed of
bright clumps of emission randomly distributed about the center, but
showing a general east to west drift (as is demonstrated in
Fig.\ref{mom1}a) which might be interpreted as rotation, or at least
coherent streaming motions. What is the dynamical state of this cloud
system and is it coupled to the motions of the ionized gas and stars
seen rotating in quite orderly motions near the nucleus of NGC\,985?

\citet{arribas} showed that {\em both the stellar and emission line
velocity field} in the central few arcsecs around the nucleus of
NGC\,985 is consistent with an orderly rotating disk , with a
kinematic axis along a PA of -45 degrees (this is the same positional
angle that connects the two nuclei). It is already clear from the
isovelocity map of Fig.\ref{mom1}a that the east-west drift seen in
the CO velocity field {\em does not} share the same kinematic major
axis as the ionized gas. Indeed, if we ignore the obvious
irregularities in the CO velocity field, and naively interpret
Fig.\ref{mom1}a as indicative of a rotating disk, it would have a
kinematic major axis of approximately -90 degrees--quite different
from the kinematic major axis of the more-uniform ionized disk.

\begin{figure}
\plotone{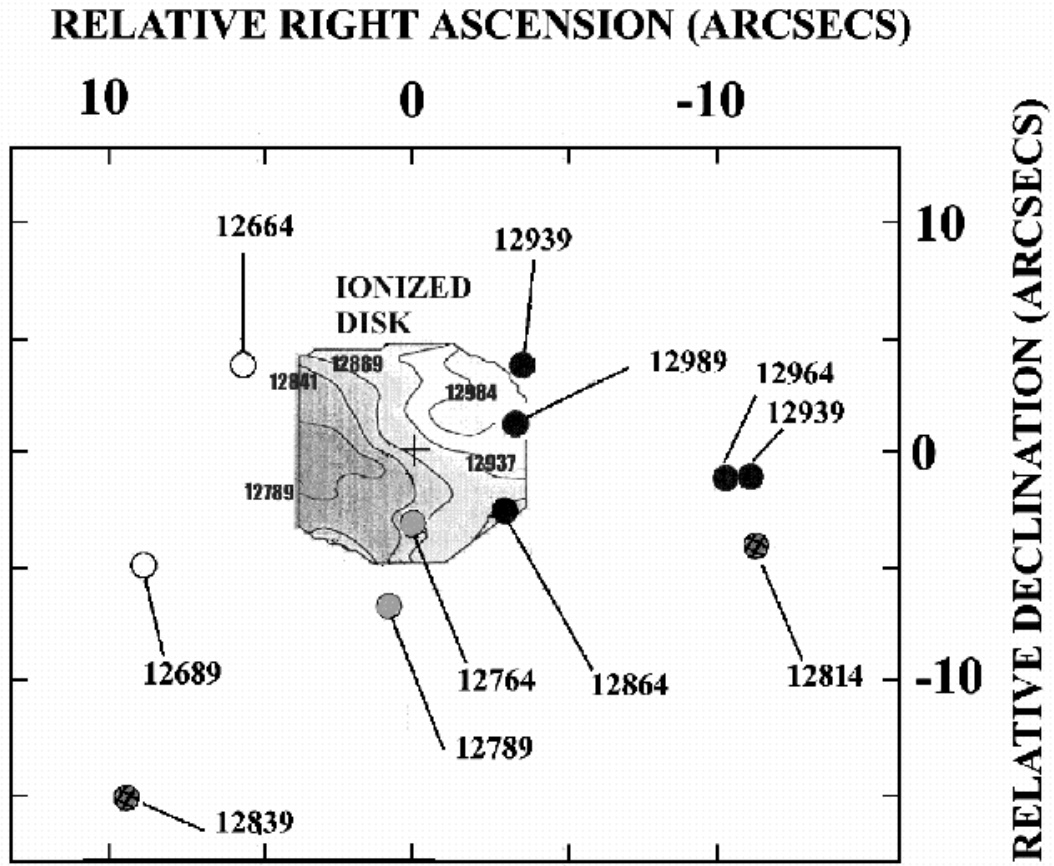}
\figurenum{11} 
\caption{
Emission centroid positions and velocities for eleven SGMCs detected
in CO, superimposed on the isovelocity map of the {\em ionized} gas
disk mapped via optical emission lines by \citet{arribas}. Note the
east-west drift of the CO centroids with increasing velocity and the
miss-match with the velocity field of the ionized gas disk (see
Section~\ref{dynamics}).\label{ionized}}
\end{figure}

We can explore the possible connections between the CO emission and
the ionized disk more clearly in Fig.\ref{ionized}, which shows the
position of the CO emission centroids of eleven of the centrally
concentrated SGMCs (and their velocities) superimposed on the
isovelocity ``spider'' diagram of the {\em ionized} gas from the work of
\citet{arribas}. This figure shows the east-west drift in the SGMC
positions with increasing velocity, but confirms that the clouds are
definitely {\em not} co-rotating with the ionized gas disk. For
example, if the CO clouds were simply clumps of molecular material
moving in circular orbits with the ionized gas, then each CO centroid
should lie on a corresponding isovelocity line as defined by the
velocity field of the ionized gas. In contrast to this, most of the CO
centroids lie at velocities which are discrepant from the velocity of
the ionized gas at the same position. A number of the CO centroids lie
outside the region where the ionized gas velocity field was measured,
but even in those cases, the velocities corresponding to those CO
centroids are inconsistent with any simple extension of the ``spider''
diagram of the ionized gas. The CO clouds are most discrepant to the
east and south of the nucleus, where they have velocities which are
approximately 100 km/s too low for the corresponding ionized gas
velocity. Only one SGMC (at V = 12989 km/s) has the same velocity as
the ionized gas at that same position (it may be coincidental-but this
is close to the position of the second (non-active) nucleus). The CO
emission, while exhibiting the same general trend of east-west motion,
appears only loosely coupled to the ionized disk, and shows some
highly discrepant velocities, especially to the south and east.

\subsection{An Unstable Collapsing Disk or Clouds Dominated by Random Motions?}

In this section we will explore two different dynamical states for the
inner CO cloud distribution. The first will assume that the east-west
drift seen in the CO centroids is due to rotation, and we will show
that these rotational velocities are insufficient to stop the CO from
sinking towards the center when compared with the likely mass in old
stars present in NGC\,985. A second alternative also exists, that the
clouds travel on non-circular orbits throughout the center-allowing
for the possibility that they are marginally stable. The latter, if
true, implies that the central CO distribution is highly disturbed,
perhaps a result of interactions between the CO clouds and an AGN
wind--an extra complication that must be taken into account when
modeling the evolution of gas in merging gas-rich systems.

If we were to interpret the overall east-west trend of velocity with
position across the center of NGC\,985 in the CO clouds as being
rotation, we can ask what mass would be required to keep the clouds in
stable circular motion.  Taking the CO centroids both east and west of
the nucleus to be representative of the maximum velocity spread
(275\,km\,s$^{-1}$), this implies that the mass needed keep the clouds
in circular orbits at a radius of r=10 arcsecs (8.25 kpc) would be
$2.89\times10^{10}$ (cosec i)$^2$\,M$_{\sun}$, where $i$ is the
inclination of the disk. Assuming an average inclination $i$ = 45
degrees, this would imply a total mass within 8.25\,kpc needed to
produce circular orbits of M$_{\rm circ}$ = $6\times10^{10}$
M$_{\sun}$. We will show, in what follows, that this mass is
unrealistically small, implying that if the clouds are in rotation,
they have insufficient velocity to remain in circular orbits, and are
likely to be falling inwards towards the center, a result that has
implications for the future evolution of NGC\,985.

Before we estimate the minimum gravitational mass in the inner region
of NGC\,985 using the flux in K-band light, we will make another
estimate of the mass required for an equilibrium configuration for the
CO clouds which does not make explicit assumptions about the shape of
the orbits. Such an approach is justified if we believe (as concluded
above) that the CO clouds may not be in circular orbits. The method,
called the projected mass estimator, was devised by \citet{bah81} for
the problem of the mass of dark matter in small groups of
galaxies. This may be more appropriate for a system in which we have
only a small number of ``test particles'' (the SGMCs) moving in the
potential of the galaxy. The methods finds a weighted mass for a
system assuming that the ``test particles'' probing the mass are
gravitationally bound to the central mass.  Making the least
assumptions about the ellipticity of the orbits, the relevant equation
is:

\begin{equation}
{\rm M_{o}} = \frac{24}
		{\pi \rm G N} \sum_{i=1}^{\rm N} v^{2}_{zi} \rm R_{i} 
\end{equation}

where $v_{zi}$ is the difference in the recessional velocity between
the i$^{th}$ SGMC, and the central massive object. R$_{i}$ is its
projected distance from the center, and N is the total number of
SGMCs.

Putting in the values for the randomly distributed SGMCs in the above
equation, and assuming the velocity and position of the AGN are the
barycentric coordinates, the total projected mass is found to be
M$_{\rm o}$\,=\,1.43$\times$10$^{11}$ M$_{\sun}$. This is more than
twice the mass found assuming circular orbits, but uses all the cloud
velocities rather than just those at the velocity extremes of the
distribution, as in the circular velocity case.

How do these dynamically estimated masses compare with the mass in
stars in the inner regions of NGC\,985? We can determine a lower limit
to the mass of stars by estimating the mass associated by old stars
via the relationship developed by \citet{thronson}. This approach uses
the K-band luminosity of a galactic region (they applied it to the
center of M51) calibrated via measurements of the mass in old stars in
the solar vicinity.  Normally such a calculation could not be applied
to a galaxy dominated by a powerful AGN, but \citet{AM} showed that at
2.2\,$\mu$m, the bright nuclear source is only a minor component of
the K-band light-which is dominated by old stars in the bulge. We are
therefore able to remove the contribution of the AGN from our
calculation. We emphasize that this is likely to be an underestimate
of the true mass, since it neglects younger stars and the possible
contribution of dark matter.

Using the maps and photometry of \citet{AM}, we are able to conclude
that within a diameter of 22 arcsecs (18\,kpc) the total K-band flux
is 33\,mJy, of which only 13\,mJy (39\%) originates within 2$\times$2
arcsecs of the AGN. Hence the total K-band flux from the region of the
random SGMC component is approximately 20\,mJy. This corresponds to a
total mass in old stars of M$_{\rm old
stars}$\,=\,1.5$\times$10$^{11}$\,M$_{\sun}$. We consider this a
conservative lower limit to the true mass.

This lower limit to the stellar mass in the inner regions of NGC\,985
is larger than the mass estimated from the velocity and positions of
the CO clouds assuming reasonable circular orbits, M$_{\rm old
stars}$/M$_{\rm circ}$ $>$ 2.5, and is closer (but still larger than)
the mass found assuming the clouds travel on non-circular orbits. This
implies that either the gas is spiraling into the center on a
relatively rapid time-scale, or that the gas clouds are exhibiting
high peculiar motions-perhaps some infalling and some expanding
outwards from the center. In both cases, this argument, when combined
with the obvious discrepancies with the velocities of the regularly
rotating {\em ionized} gas disk suggests that the CO clouds are in a
highly disturbed state and are probably not in equilibrium with the
gravitational potential.

\section{Nuclear Asymmetries - Further Evidence from ISO}\label{sect_asymetry}

\begin{figure} 
\plotone{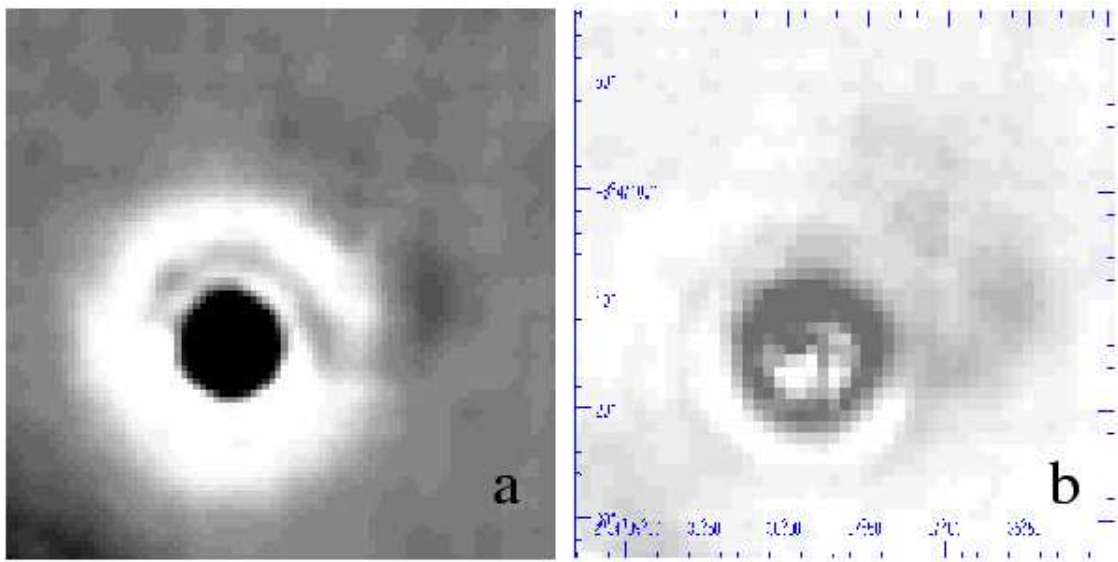}
\figurenum{12} 
\caption{a) 
The 15\,$\mu$m image of NGC\,985 after the unsharp mask
technique has been applied. The asymmetric emission and a bright
``ridge'' to the north and west of the nucleus. b) 15\,$\mu$m image after 
a PSF fit and subtraction of the Seyfert
nucleus. The field of view is the same as in a). Even though less well
defined than the unsharp image one can easily see the same extend in
MIR emission to the north of the nucleus.\label{shell}}
\end{figure}

The CO distribution and kinematics are not the only features of this
galaxy which suggest that the inner galaxy is in a non-equilibrium
state. In Figure~\ref{shell}a we show that the MIR emission near the
nucleus is not symmetrical, but that there is a bright arc or shell to
the north and west of the nucleus at an angular distance of about 5
arcsecs \citep[see also][]{iso_rings}. The figure shows the
application of a simple ``unsharp mask'' technique to the $\lambda$15$\mu$m ISOCAM
image. This filtering technique is sensitive to small-scale structure
in the image on the scale of a few arcsecs.

To further explore this feature, which is also obvious on
independent images taken at other wavelengths (the 9.62$\mu$m and 11.4$\mu$m
images), we performed a series of experiments using the theoretical
monochromatic ISOCAM PSF for $\lambda$15$\mu$m, smoothed appropriately
to take into account the width of the filter. The PSF was also rotated
to match exactly the micro-scan direction of the observations.

We show in Figure~\ref{shell}b the result of an iterative removal of
the PSF from the peak of the Seyfert nucleus. The image shows the
result of three iterations of removal, in which the PSF was subtracted
from the peak in the map with a loop-gain of 0.5 \citep[similar to the
technique CLEAN of][ used in radio astronomy]{hogbom}. The following
conclusions were drawn from our experiments: a) The nuclear source
itself is slightly extended on the scale of a few arcsecs and, b)
there is an excess of emission to the north and west of the nucleus
which cannot be explained by the properties of the detector or the
observing technique.  Furthermore, these bright regions fall on the
null of the first Airy ring of the optical system -- again an
indication that the emission is real.

In conclusion, we believe that the ISO images show, on several
independent filters, evidence of an asymmetry in the dust distribution
near the nucleus which we are unable to explain as an instrumental
artifact. The emission may be shell-like, but because the emission
falls so close to the bright nucleus we would prefer to await higher
resolution observations before making a definitive statements about
its morphology. However, the emission does support the view that the
dust and gas are highly disturbed and not always directly coupled,
since this asymmetric dust feature is {\em not} seen in the same
direction as the main CO distribution (although, as discussed in
Section~\ref{sect_kinematics}, there may be a faint CO counterpart
seen in one channel).

\section{Mid-IR Spectral Energy Distribution}\label{sect_mirsed}

It has been shown \citep{lutz,diagnostic} that a galaxy hosting a
dominant AGN is clearly different in the MIR from one dominated by a
nuclear starburst because the AGN heats the dusty torus surrounding
it. Perhaps because of grain destruction in the powerful AGN radiation
field, the well known emission bands at 6.2, 7.7, 8.6, 11.3 and
12.7\,$\mu$m (usually attributed to Polycyclic Aromatic Hydrocarbons)
are extremely weak or completely missing from the MIR spectrum of
powerful AGNs, and this can provide an additional diagnostic if the
MIR extinction of the nucleus is not too great.

To examine the MIR characteristics of the nucleus of NGC\,985, we
display in Figure~\ref{spectra}b the four MIR flux densities measured
over an aperture 12 arcsecs in diameter centered on the radio
continuum peak (from Table~\ref{tbl_mir}).  This aperture covers the
majority of ISO emission associated with inner regions of NGC\,985.
The width of the corresponding ISOCAM broad-band filters is also
displayed with the horizontal lines. For comparison, we also show the
MIR spectra of two very different emission regions, an AGN dominated
MIR spectrum from the nucleus of NGC\,1068 \citep[see Fig.~4
of][]{emeric}, and a typical star formation-dominated MIR spectrum
from knot B of the Antennae galaxies
\citep[see][]{vigroux,mirabel}. Both spectra have been redshifted to
the same rest wavelength as NGC\,985 and their flux was normalized to
the 11.4\,$\mu$m flux of NGC\,985.

An inspection of the SED in Figure~\ref{spectra}b suggests that the
longer-wavelength spectrum of the nucleus of NGC\,985 looks more like
a nucleus dominated by star formation, than one where the dust is
heated directly by an AGN, since the spectrum of NGC\,985 does not
rise as fast as that of NGC\,1068, especially in the 10--15\,$\mu$m
range. Only at $\lambda$4.5\,$\mu$m is there a hint of a
``warm bump'' from the slightly elevated flux at that
wavelength. However, since this could be due to contribution from old
stars of the bulge and we lack more definitive diagnostics in the
5--10\,$\mu$m range \citep{diagnostic,emeric}, further observations
would be necessary to confirm this possible result.

It is of interest to note that our first-order deconvolution of the
nuclear emission from the AGN (see Section~\ref{sect_asymetry})
suggested that the nuclear emission maybe slightly extended relative
to the theoretical PSF for the telescope. If so, we tentatively
associate this emission with the stellar and hot-ionized gas disk
which is seen in the central few arcsecs of the galaxy by
\citet{arribas} from optical observations. This inner rotating
``disk'' must be composed of both stars and ionized gas since both are
seen to co-rotate along a line joining the two embedded galactic
nuclei. The MIR observations may be pinpointing the formation of a
proto-disk at the core of what will eventually become a collisional
remnant. Could this be direct evidence for the formation of disky
cores in elliptical galaxies? Higher resolution MIR observations
should allow us to determine whether the MIR emission does indeed
coexist with the inner rotating ionized disk seen by \citet{arribas}.

\section{Non-nuclear Star formation sites in NGC\,985}\label{sect_srf}

We had seen from the CO, radio, and MIR maps that NGC\,985 contains a
bright ridge of stars, gas and dust that extends from the double
nucleus all the way out to the stellar ring to the west. Furthermore,
in the paper by \citet{RES} it was shown that the entire length of the
optical/IR ridge is populated with powerful \ion{H}{2} regions, the
most powerful of which lie towards the end of the ridge, and in two
additional \ion{H}{2} regions in the ring. Although the ISO
observations show MIR emission associated with the bright H$\alpha$
emitting regions, we note that the MIR emission does not always mimic
the brightness variations from the optical line-emitting regions. For
example, the two bright \ion{H}{2} regions in the outer ring
\citep[knot A and B in][]{RES} are much stronger in the ISO
observations than the end of the linear structure \citep[knots G \& H
of][]{RES}, despite the fact that both sets of regions have very
similar observed H$\alpha$ luminosities.  Such differences may relate
to differences in the relative populations of small and large dust
grains in the various regions \citep[as suggested by][ to explain
differences in the MIR properties of the supergiant CO complexes found
associated with star forming regions in the Antennae
galaxies]{wilson}, or they could simply reflect different internal
optical extinction within the optical nebulae.

One explanation for the high star formation rates seen along the
linear ridge is suggested by the high velocity dispersion in the CO
gas in this region (see discussion in
Section~\ref{sect_evolution}). For example,
\citet{stanford,wilson,gao} have shown that in NGC\,4038/4039, one
of the most powerful star formation sites corresponds to a region of
high CO velocity dispersion \citep[the region called SGMC~4--5
by][]{wilson}. In our case, we observe a similar phenomenon. In
particular, the region towards the end of the linear structure
corresponds to the highest velocity dispersion seen in the molecular
gas (150\,km\,s$^{-1}$). This region also contains some of the
brightest star formation sites seen along the optical emission ridge,
since knots G and H from \citet{RES} have combined luminosities in the
H$\alpha$ line of 7.1$\times$10$^{40}$\,erg\,s$^{-1}$, a value
approximately 15 times the luminosity of the 30 Doradus nebula in the
LMC.

Even the outer ring of NGC\,985 has some odd characteristics. For
example, we note from the overlays in Figure~\ref{spectra}a, that
although the MIR emission at $\lambda$15$\mu$m falls almost on top of
the contours of bright ring \ion{H}{2} regions A and B of \citet{RES},
the CO emission (see Figure~\ref{comap}) falls significantly {\em
inside} the ring, suggesting that the CO near the \ion{H}{2} regions
themselves has been destroyed, but the dust grains associated with the
\ion{H}{2} regions, clearly seen in the ISO maps, have not! This
underlines the complex relationship between CO emission and dust
emissivity in colliding galaxies -- a point already made cogently by
\citet{wilson} for the NGC\,4038/39 system.

To further complicate the star-formation story for NGC\,985, we note
that the radio continuum emission from the ring also fails to
correlate spatially with either the optical, radio or the CO emission!
The main $\lambda$3.5\,cm emission regions, seen in
Figure~\ref{radiomap}, lies inside the bend in the ring, and the
offset is significant. This offset between the radio continuum and the
current star formation sites has been seen before in ring galaxies
\citep[e.g.][]{7zw_iso} and may be the natural result of the
propagation of a star formation wave through the outer disk of the
target galaxy. The radio continuum emission, if created by cosmic rays
from supernova explosions, will tend to lag behind the regions of
current star formation site because time must elapse for the massive
stars to explode.

We can attempt to estimate the number of supernova needed to create
the observed radio continuum emission. Assuming that the the radio
emission is predominantly non-thermal (the spectral index is -0.6
which suggests mix of thermal and non-thermal emission) then we can
estimate the supernova rate from equation 8 of \citet{cy} :

\begin{equation}
\rm L_{\rm NT}(\rm W\,\rm Hz^{-1}) =
	1.3\times10^{23} (\frac{\nu}{1\,\rm GHz})^{-\alpha}
                      ( \frac{ \rm R_{\rm SN}}{ \rm 1\,yr})
\end{equation}

where L$_{\rm NT}$ is the non-thermal radio luminosity, $\alpha$ is
the radio spectral index, and R$_{\rm SN}$ is the rate of type II
supernova. For the western radio component, the radio luminosity at
6\,cm, L(6cm)~=~3.3$\times$10$^{21}$ W Hz$^{-1}$, which corresponds to
a maximum (assuming all non-thermal emission) SN rate R$_{\rm
SN}$\,=\,0.067\,yr$^{-1}$. Is such a supernova rate reasonable, given
the know rate of star formation in the giant \ion{H}{2} regions near
the radio component?

To test this, we can calculate the rate of type II supernova that
would be expected from knots A and B combined (the adjacent knots to
the radio emission region), based on the star formation rates in those
\ion{H}{2} regions. If we assume a Salpeter IMF and a reasonable
cutoff to the upper mass limit on the IMF we reverse equation 12 of
\citet{cy} to:

\begin{equation}
\rm R_{\rm SN} = \rm L_{\rm H\alpha}(\rm W)/(1.1\times10^{36}) \rm yr^{-1}
\end{equation}

Using the combined values for knots A and B from \cite{RES} of
L(H$\alpha$) = 9.0$\times$10$^{33}$ W, we find R$_{\rm SN}$=0.01, a
factor of six times too low to explain the observed radio emission.
However, when correcting for an internal extinction of 2 magnitudes
based on optical spectra of the knots \citep{B98} these rates come
into reasonable agreement. We conclude that a burst of star formation
in the past, with a magnitude similar to that currently occuring
further out in the ring is quite capable of supplying the required
cosmic rays needed to power the radio source.

\section{UV Absorbers associated with CO emitting clouds: Coincidence
or Physical Connection?}\label{sect_uvabs}

Several strong absorption features are seen in the blue wings of the
broad Ly$\alpha$ and \ion{N}{5} 1238/1242 (S. Penton \& J. Stocke
private communication). Intrinsic lines of this kind are found in
roughly 15\% of Seyfert galaxies, and seem to represent a population
of warm clouds expelled in a nuclear wind. Independent evidence for
warm absorbers in NGC\,985 are found from the X-ray analysis of
\citet{Brandt}. The HST spectrum kindly shared with us  by S. Penton and
J. Stocke show six absorption lines seen in
Ly$\alpha$. All except the most blueshifted line is also seen in
absorption against \ion{N}{5}. The velocities for the lines are listed
in Table~\ref{tbl_uv}. The highest velocity line, around
13,065\,km\,s$^{-1}$ in Ly$\alpha$, is narrow and isolated (FWHM of
approximately 50\,km\,s$^{-1}$) whereas the other lines, although also
narrow, may be part of a broad blended system extending over a full
width of 600\,km\,s$^{-1}$. Nevertheless, these individual narrow
features are (except the most blueshifted one) seen in absorption
against \ion{N}{5}$\lambda$1239,1243 as well.

Interestingly, we find that the two most red-shifted lines correspond
to velocities at which CO gas is seen projected against the
nucleus. In fact, from Fig.~\ref{ukirt} it can be seen that, because
of the off-set in the main CO distribution, gas is seen against the
nucleus in only three regimes -- between V=12664--12689,
V=12,764--12789 and V=12989--13039\,km\,s$^{-1}$. The highest and
lowest velocity CO emission feature seen against the nucleus
correspond quite well to the velocities of the warm absorbers.

The high velocity absorption line (and the CO in emission) lies close
to the systemic velocity of the non-active nucleus N2. The narrow
width of the UV absorber implies that it might be ISM material
associated with N2 (despite this being seen in absorption against N1).

The UV absorber system observed at V= 12,670\,km\,s$^{-1}$ also seems
to correspond to the velocity of gas seen in CO emission in front of
the nucleus -- in fact it corresponds to CO seen in two adjacent
channels (spread over 50km\,$^{-1}$). These emission features form part
of the randomly-distributed cloud population discussed earlier.  Could
the UV absorber and the CO emission be related -- perhaps through an
interaction of the AGN wind with the infalling clouds? One argument
against this is that the CO clouds are cold, and yet the UV
absorption-line clouds are warm. However, if an infalling cloud was,
through the chaotic nature of the infall, to encounter an AGN-driven
wind, it may become heated and its outer envelope may well become
ionized. Although our CO observations cover two more of the
uv-absorption line velocities, no emission was detected from the more
blue-shifted components.  Perhaps dense molecular clouds encountering
the wind do not survive long, and are evaporated rapidly as they
become entrained in the wind. We note that this may be the first time
there had been a possible identification between intrinsic Ly$\alpha$
absorption line components and CO emission, although CO absorption
lines have been discovered associated with intervening galaxies in the
direction of millimeter bright high-z sources \citep[e.g.][]{wiklind}.

\section{NGC\,985: Present and Future Prospects}\label{sect_evolution}

The presence of a faint extended (r = 10\,kpc) and probably unstable
population of SGMCs in the inner regions of NGC\,985 is generally
unusual for merging or collisional systems, although it does share
some similarities with some well studied cases. For example, in the
case of Arp\,220 \citep{Sco97}, the CO, although highly concentrated
towards the center, involves two embedded nuclei revolving within a
(CO rich) rotating disk. Similarly, the peculiar ring galaxy Arp\,143
\citep{higdon97}, like NGC\,985, also contains two molecular
components, one concentrated in the ring, and another concentrated in
the nucleus. Both these systems have highly centralized CO
distributions in rapid rotation and may suggest one possible direction
in which NGC\,985 is evolving. Alternatively, as we shall discuss
below, NGC\,985 may actually have already been through the dense
centrally-concentrated stage (ULIRG stage--\citet{sanders96}), and its
diffuse clumpy CO distribution may represent the early stages of the
disruption of the central disk by a combination of AGN and
star-formation-driven winds from the center. Both possibilities could
explain the present state of NGC\,985.

\subsection{NGC\,985: The Formation or Destruction of a ULIRG?}

Despite an incomplete knowledge of the details of the interaction, the
observations presented here provide a compelling case for a bright
future for NGC\,985! The peculiar CO distribution, decoupled as it is
from the inner rapidly rotating ionized and stellar disk, suggests the
gas is highly disturbed, and some of it may be either falling towards
the center (see Section~\ref{sect_kinematics}), or may be in
marginally stable, but non-circular orbits. There are however many
questions that are raised by the observations. What will be the fate
of the 2$\times$10$^{10}$\,M$_{\sun}$ of molecular gas over time? What
is the significance of the intrinsic warm absorbers seen in the UV and
X-ray spectra of the Seyfert nucleus and their relationship to the
random component of the CO clouds? Finally, is NGC\,985 likely to
become much more IR-luminous in the future, evolving towards a ULIRG
state, or is there evidence that it has already passed through that
state and that its activity may well be in decline?
 
The current observations provide only partial answers to these
questions. Firstly, in the absence of a disturbing nuclear wind, the
CO reservoir cannot remain out of equilibrium for more than a few
crossing times, given the dissipative nature of gas. Taking the
overall scale of the CO structure (about 20--25 arcsecs = 16--20\,kpc)
and dividing by the total velocity-spread in the CO
(Fig.~\ref{cospectrum}) of 400\,km\,s$^{-1}$, provides a rough
timescale for the dissipative collapse of the gas of 45--50\,Myrs. If
the gas is indeed infalling and we conservatively assume that 2 such
periods would be needed to allow the gas to settle, then we might
expect the large-scale accretion of the gas into the center to occur
over the next 100\,Myrs. If this was a smooth process, this would
provide an average accretion rate of 200\,M$_{\sun}$ yr$^{-1}$. Given
the peculiar distribution of the gas, it is unlikely to be a smooth
event. If each of the randomly distributed SGMCs falls into the center
separately they will each induce a period of sporadic activity as the
gas is assimilated into the central disk. It is hard to believe that
under these condition, the nuclear activity in NGC\,985 will not
increase significantly -- perhaps further increasing the already
luminous nature of the AGN.

What might be the effect of this infall on an already active nucleus?
One possibility is that some infalling gas will become entrained in a
wind presently blowing from the Seyfert core. There is evidence from
the X-ray spectrum of a significant soft X-ray component to
NGC\,985. Some of this may relate to the nuclear star formation which
we believe must be present in NGC\,985 from the ISO spectral energy
distribution, but some may be due to an outflowing wind. It is
possible that CO clouds raining towards the center become entrained
and accelerated in the outflowing wind and as a result become
heated--creating warm outer atmospheres which we tentatively associate
with the UV absorption lines and X-ray absorbers. Indeed it would seem
natural that inflowing material in tidal interactions involving active
nuclei would inevitably encounter winds from the AGN as some
stage. This may explain the fact that intrinsic UV absorbers are seen
only in a sub-set of cases in UV spectra of Seyfert galaxies (the
major accretion stage?). Further observations would be needed to see
if the appearance of UV absorbers in AGN spectra is correlated with
the appearance of highly asymmetric CO distributions, like that of
NGC\,985. In any event, our present observations provide for the
possibility that the tidal fueling of AGNs must take into account the
possibility of potentially complex interactions between the infalling
gas and the turn-on of an AGN wind.

The possibility that a nuclear wind could interact, or even disrupt
the formation of a central nuclear molecular disk leads to another
possible path for the evolution of NGC\,985. Could the SGMCs seen in
the inner region represent the ending of the ULIRG stage rather than
the beginning of it? We have mentioned that many interacting systems
show highly concentrated CO disks, and NGC\,985 is quite exceptional in
that its CO distribution is highly asymmetric-indeed the brighter CO
is seen away from the nucleus in the outer disk region, and the CO in
the clouds with highly peculiar motions are significantly
fainter. Could these clouds represent the disrupted remnants of a much
more centrally concentrated CO distribution--the stage that is assumed
to be associated with ULIRGs?

The question of whether NGC\,985 about to {\rm become} a ULIRG, or
alternatively has recently passed through that stage can be tested
with future observations.  Observations of the elliptical-like bulge
component which surrounds the double-nucleus may help to determine
whether the galaxy has recently experienced a major episode of global
distributed star formation-a symptom of an ongoing or post-merger
history \citep{schweizer}. Also, if NGC\,985 is emerging, rather than
entering a ULIRG stage, it would be expected to show a very extensive
soft X-ray emission halo resulting from a recent violent star
formation history. High-resolution multi-wavelength IR observations of
the bulge may also help to determine what contribution the bulge makes
to the strong extended mid-IR emission seen associated with
NGC\,985. Such observations would help to determine whether the
R$^{1/4}$--law bulge itself contributes a significant amount to the
total MIR energy budget of the NGC\,985 system.

\section{Conclusions}

Observations of NGC\,985 with ISO, the VLA and BIMA have led to the
following conclusions about this system of galaxies:

1) A MIR luminous AGN is centered on the brighter component of a
double nucleus. The MIR nucleus, which is detected in all 4 ISOCAM
bands (from 4.5\,$\mu$m to 15\,$\mu$m) is consistent with emission from a
rapidly rising thermal continuum. The slope of the continuum is
consistent with dust heated predominantly by star formation, rather
than the AGN itself. We tentatively associate this slightly-extended warm
nuclear dust emission with the rapidly rotating stellar and gaseous
disk seen at optical wavelengths by \citet{arribas}. This disk, which
rotates along the line joining the two embedded galactic nuclei, may
be forming as a consequence of the merger, resulting from inflowing
gas to the nucleus.

2) Our BIMA observations of NGC\,985 support the view that it is in a
violent state of merger. The CO emission, though covering the nucleus,
is distributed in a highly asymmetric distribution, with the peak
line-of-sight-column significantly offset from the compact
double-nucleus. This distribution resembles the molecular distribution
of a number of quasars recently mapped in CO, many of which have
asymmetric and irregular CO distributions \citep{Evans01}.  Most of
the cool gas in NGC\,985 is contained within this large molecular
reservoir, which, when adopting standard conversion factors from CO to
H$_2$ masses, contains 1.8$\times$10$^{10}$\,M$_\sun$.

3) Although the highest CO column densities are seen away from the
nucleus, we isolate a population of SGMCs which may be mildly
collapsing towards the center on a timescale of 60--100\,Myrs. These
clouds, which extend within 9--10\,kpc of the AGN, were identified
through their random distribution and unusual motions which are
different from the more coherent motions of material further out in
the CO ridge. This two-component nature of NGC\,985, may indicate a
transition time in a merging system in which the motions of the clouds
begin to loose their original ordered motions in the collision, and
begin to rain down on the center through dissipation. Sporadic
accretion rates are likely to exceed 100\,M$_\sun$\,yr$^{-1}$ if all
the molecular material present fall to the center in a few dynamical
crossing times.

4) NGC\,985 currently has a low ratio of L$_{\rm IR}$/M(H$_2$) =
10. This value lies in the range found for low-luminosity starburst
galaxies, despite its abnormal CO distribution and large, as yet
mainly untapped, molecular reservoir. This low value may indicate that
either i) NGC\,985 is about to enter the ULIRG stage as the huge
molecular reservoir accumulates at the center, or ii) that the galaxy
is {\rm emerging from} the ULIRG phase, and that the action of a
powerful AGN or nuclear starburst wind is inhibiting star formation in
the inner regions.

5) We tentatively associate two CO emission features in the direction
of the AGN with two out of six narrow UV absorption-lines seen in an
HST study of NGC\,985. One of these CO emission features is associated
with the population of clouds that might be mildly infalling towards
the center, and this raises the possibility that the infalling clouds
may sometimes become entrained in an outflowing AGN wind. How such
interactions may affect the fueling of an AGN over time is not
clear. However, these interactions may play an important role in the
disruption of centrally concentrated molecular disks in starburst
systems that contain powerful AGN.

6) The CO furthest away from the center shows more ordered motions and
is associated with the optical/IR ridge-probably a spiral arm or
ring-segment containing compressed gas. The tip of the CO ridge, the
most distant from the double-nuclei, shows the highest velocity
dispersion -- a property that seems to correlate with a high rate of
local star formation.

7) The ISO observations also reveal warm dust associated with the
western optical/IR ridge extending from NGC\,985. The dust is
consistent with grains heated by bursts of massive star formation
which are observed along the ridge and into the outer optical ring of
NGC\,985. As in the case of the Antennae galaxies, the MIR dust emission
does not always scale linearly with other observed properties of the
star formation regions (e.g. H$\alpha$ luminosity, molecular cloud
mass). We find that in one region, MIR emission is seen associated
with a bright optical \ion{H}{2} complex, but no $^{12}$CO(1-0)
emission is observed at that position, indicating a complicated
relationship between dust grain lifetimes and molecular gas lifetimes
in massive star formation regions. Our radio continuum observations
also provide evidence of a previous episode of star formation which
occurred in the recent past inside the ring.

\acknowledgments
This work was supported in part by NASA grant NAG-5-3317. The authors
thank S. Penton and J. Stocke (U. of Colorado) for communicating the
results of their HST/STIS study of NGC\,985, and for kindly allowing
us to present the absorption-line velocities in Table~\ref{tbl_uv} in
advance of their publication of the results. We also appreciate the
help of A. M. P\'erez Garc\'{\i}a and J. M. Rodr\'{\i}guez Espinosa
(IAC, Spain) for providing their deep R-band image of
NGC\,985. P. Appleton would like to thank V. Charmandaris and
I. F. Mirabel for their hospitality during the ISO reduction stage of
the analysis in France.  V. Charmandaris would also like to thank
O. Laurent (Saclay \& MPE) for help and suggestions on the optimum
analysis of ISOCAM data. The authors wish to thank an anonymous
referee for helpful comments on the original manuscript.

\clearpage






\clearpage

\begin{deluxetable}{lcccccc}
\tablecaption{Radio and Mid-Infrared Properties of NGC\,985\label{tbl_mir}}
\tablewidth{0pc}
\startdata \\
\tableline 
\tableline \\
  &ISO-LW1 & ISO-LW7 & ISO-LW8 &ISO-LW3 & VLA&VLA \\

Region & S(4.5$\mu$m) & S(9.7$\mu$m) & S(11.4$\mu$m) &
S(15$\mu$m)  & S(3.5\,cm) & S(6\,cm) \\
 & mJy&mJy&mJy&mJy&mJy&mJy \\
\tableline \\
Entire Galaxy & 55.2 & 113.5 & 154.4 & 168.9 & 2.28 & 4.47  \\
Nucleus/Bulge\tablenotemark{a} & 46.5 & 105.6 & 142.5 & 163.6 & 1.49 & 3.38\tablenotemark{b}  \\
Eastern SF Knot/ring\tablenotemark{c} & 0.8  & 5.4 & 6.2 & 3.5 & 0.79& 1.09 \\


\tablenotetext{a}{Flux measured in a circular aperture of radius 12
arcsecs centered on the nucleus.}
\tablenotetext{b}{\citet{ulv84} quote a 6\,cm flux for the nucleus of
2.5\,mJy with the VLA in A configuration, but mention that this is
probably an underestimate of the flux since they note that the flux
they detected varied when the array was tapered differently. Our
observations, which were made with many short spacings in the D
configuration, confirm that the A-array observations missed a
substantial fraction of the flux.}
\tablenotetext{a}{The ISO fluxes were evaluated over the bright MIR emission knot centered at $\alpha$(J2000) = 2h 34m 36.5s, $\delta$(J2000) = -8 d 47' 10'', whereas the radio emission was evaluated over the larger area defined by the radio maps of the second western source seen in Fig. 3.}
\enddata
\end{deluxetable}

\newpage 

\clearpage

\begin{deluxetable}{lcl}
\tablecaption{Global properties of NGC\,985\label{tbl_global}}
\tablewidth{0pc}
\startdata \\
\tableline 
\tableline \\
 & & Notes \\
\tableline \\
Assumed Distance (Mpc)          &  170          &   \\
m$_{\rm v}$~(mag)               &  13.57       & 
                                : from \citet{opt_rings}.\\
L$_{\rm IR}$~(L$_\sun$)         &  1.8$\times$10$^{11}$ & 
                                : definition of \citet{sanders96}.\\
M(\ion{H}{1})~(M$_\sun$)        &  $<$\,3.6$\times$10$^9$  & 
                                : from \citet{hec}\tablenotemark{a} \\
S$_{\rm CO}$=$\int\,Sdv$~(Jy\,km\,s$^{-1}$) 	& 54   &
				: from \cite{cathy}, filled aperture\\
S$_{\rm CO}$=$\int\,Sdv$~(Jy\,km\,s$^{-1}$)     & 54.8 &
                                : (BIMA) this work.\\
CO Line Width (FWHM, km\,s$^{-1}$)    & 408 $\pm$ 15 &
				: this work.\\
M(H$_2$)~(M$_\sun$)             &  1.8$\times$10$^{10}$ & 
                                : this work\tablenotemark{b}.\\
L$_{\rm IR}$/M(H$_2$)~(L$_\sun$/M$_\sun$)       & 10    & \\
M(H$_2$)/M(\ion{H}{1})          & $>$\,5.5        & \\
\enddata

\tablenotetext{a}{Based on \citet{hec} after correcting for our adopted
value of H$_0$.}
\tablenotetext{b}{Using a N(H$_2$)/I$_{\rm CO}$ =
3$\times$10$^{20}$\,mol
\,cm$^{-2}$ (K\,km\,s$^{-1}$)$^{-1}$}

\end{deluxetable}

\newpage 
\clearpage

\begin{deluxetable}{cc}
\tablecaption{UV (HST/STIS) observations of nuclear absorption
lines\tablenotemark{a}\label{tbl_uv}}
\tablewidth{0pc}
\startdata \\
\tableline
\tableline \\
Ly$\alpha$\tablenotemark{b}     & Inferred heliocentric cloud velocity\\
\AA             & km\,s$^{-1}$ \\
\tableline \\
1264.787 $\pm$ 0.02  & 12,113 $\pm$ 4 \\
1265.317 $\pm$ 0.02  & 12,243 $\pm$ 4 \\
1265.927 $\pm$ 0.06  & 12,394 $\pm$ 4 \\
1266.396 $\pm$ 0.05  & 12,509 $\pm$ 4 \\
1267.047 $\pm$ 0.02  & 12,670 $\pm$ 4 \\
1268.611 $\pm$ 0.00  & 13,053 $\pm$ 4 \\
\enddata

\tablenotetext{a}{Kindly provided by S. Penton \& J. Stocke.}
\tablenotetext{b}{The rest wavelength of Ly$\alpha$ is 1215.67\,\AA.}

\end{deluxetable}

\newpage

 %
%
 
\clearpage 

\end{document}